\documentclass[10pt,superscriptaddress,aps,twocolumn,pra,longbibliography,floatfix]{revtex4-2}

\usepackage{amssymb,amsmath,amsbsy,amsgen,amsfonts,amsthm,mathtools,mathrsfs,amstext,bm}\usepackage{array}
\usepackage{color}
\usepackage{float}
\usepackage[hidelinks]{hyperref}
\usepackage{graphicx,epstopdf}
\usepackage{verbatim}
\usepackage{url}
\usepackage{booktabs}
\usepackage{quantikz}
\usepackage{capt-of}
\allowdisplaybreaks[1]

\begin{document}

\title{Quantum state tomography with measurement-conditioned diffusion models}
\author{Woohyun Choi}
\affiliation{Department of Physics, Kyungpook National University, Daegu 41566, Korea}
\affiliation{Department of Physics, Hanyang University, Seoul 04763, Korea}

\author{Changsuk Noh}
\email[Contact author: ]{cnoh@knu.ac.kr}
\affiliation{Department of Physics, Kyungpook National University, Daegu 41566, Korea}

\author{Changhyoup Lee}
\email[Contact author: ]{changhyoup.lee@gmail.com}
\affiliation{Department of Physics, Hanyang University, Seoul 04763, Korea}
\affiliation{Hanyang Institute for Quantum Science and Quantum Technology, Hanyang University, Seoul 04763, Korea}
\affiliation{Research Institute for Natural Sciences, Hanyang University, Seoul 04763, Korea}

\date{\today}

\begin{abstract}
We introduce \mbox{Diff-QST}, a measurement-conditioned diffusion model for quantum state tomography that learns from an ensemble of input states and reconstructs a physical density operator from noisy, possibly incomplete Pauli measurements in two or three denoising steps. In simulations of two to five qubits, \mbox{Diff-QST} has higher mean fidelity than maximum-likelihood estimation (MLE) and a semidefinite-programming least-squares estimator (SDP) in all $120$ tested conditions at nominal noise. After training, GPU reconstruction is faster than both in condition-averaged time at every system size and, at four and five qubits, in every condition. At four and five qubits, though not at two and three, \mbox{Diff-QST} also has higher fidelity than an architecture-matched regression baseline trained without diffusion. At five qubits, over the tested numbers of observables and shots per observable, it reaches a specified fidelity with $25\%$--$67\%$ fewer total shots than this baseline. Under state and noise distributions that differ from those used in training, \mbox{Diff-QST} retains its fidelity advantage over MLE and SDP in most tested conditions, although it can reverse. \mbox{Diff-QST} is therefore suited to fast tomography from noisy, incomplete Pauli measurements of states represented by its training ensemble.
 
\end{abstract}

\maketitle

\section{Introduction}\label{sec_intro}
\raggedbottom

Quantum state tomography (QST) is a standard method for determining the state that a device has actually prepared~\cite{james2001,nielsen2010}.
The state is reconstructed from measurement statistics obtained on identically prepared copies. Because full QST estimates the entire density operator, from which the statistics of any measurement follow, it serves as a reference point for more resource-efficient characterization methods~\cite{eisert2020quantum, flammia2011direct, huang2020}.
An arbitrary $n$-qubit density operator has $4^n-1$ independent real parameters, so full reconstruction becomes increasingly demanding as the system grows~\cite{cramer2010,gross2010}. At fixed accuracy, even optimal tomography requires a number of copies that grows with the Hilbert-space dimension and the state rank~\cite{haah2017}.
On present hardware, these scaling challenges are compounded by imperfect data: expectation values are estimated from a finite number of repetitions, and measurement outcomes can be misassigned~\cite{bravyi2021,chen2023}. Decoherence can also make the prepared state mixed~\cite{nielsen2010}.

Maximum-likelihood estimation (MLE) selects the density operator that maximizes the likelihood of the measurement data~\cite{hradil1997,rehacek2001,rehacek2007}, whereas physical least-squares estimates can be obtained by projecting a linear reconstruction onto the set of density operators~\cite{smolin2012}. For the Pauli data considered here, we use a regularized weighted least-squares estimator solved by semidefinite programming, denoted SDP and defined in Sec.~\ref{sec_method_reference}.
MLE and SDP both enforce physicality and incorporate an assumed statistical and measurement model into their objectives. Both must be rerun for each new data set and become more costly as the Hilbert-space dimension grows.
A separate line of work instead reduces the number of measurements required by exploiting structure in the state, as in compressed sensing~\cite{gross2010} and tensor-network tomography~\cite{cramer2010,sunada2026}, or by estimating many properties from randomized measurements without reconstructing the full density operator~\cite{huang2020}.

\begin{figure*}[!t]
  \centering
  \includegraphics[width=\textwidth]{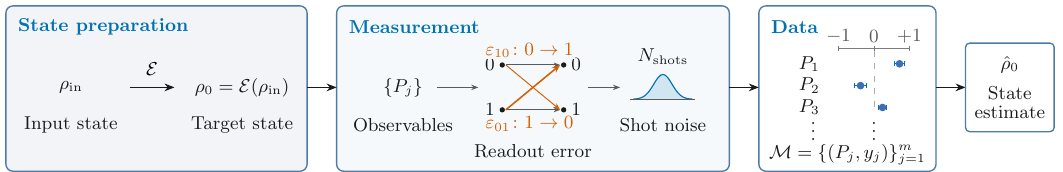}
  \caption{Overview of the state reconstruction task. The shaded blocks show the simulated preparation and measurement used to generate the data sets. Preparation noise $\mathcal E$ maps the input state $\rho_\mathrm{in}$ to the target state $\rho_0$, and a subset of $m$ Pauli observables $P_j$ is measured with asymmetric readout error (orange arrows) and shot noise. The white blocks show the resulting measurement data $\mathcal M=\{(P_j,y_j)\}_{j=1}^m$, which pair each observable with its empirical expectation value $y_j$ (error bars, one shot-noise standard deviation), and the state estimate $\hat\rho_0$ of $\rho_0$ reconstructed from $\mathcal M$ by \mbox{Diff-QST}, MLE, or SDP.}
  \label{fig:forward}
\end{figure*}

Neural-network-based QST includes per-instance methods, which fit a neural representation of the state or its measurement statistics separately to each tomographic data set~\cite{torlai2018,schmale2022,carrasquilla2019}, and methods that learn a single map for an entire ensemble of states.
Neural-shadow tomography combines per-instance fitting with an infidelity objective estimated from classical shadows~\cite{wei2024shadow}.
The second category takes several forms: supervised networks trained on simulated states map measurement data directly to density operators~\cite{lohani2020,danaci2021,koutny2022}, with reconstruction from incomplete measurements also demonstrated~\cite{lohani2020,danaci2021}; generative query networks instead predict the statistics of unperformed measurements from any number of measurement-outcome pairs~\cite{zhu2022}; neural adaptive tomography replaces the Bayesian update of adaptive tomography with a trained network~\cite{quek2021}; and supervised neural filtering mitigates state-preparation-and-measurement errors before reconstruction~\cite{palmieri2020}. Conditional generative adversarial tomography also supports direct reconstruction after pretraining on related states~\cite{ahmed2021}. More recent quantum-aware transformers encode measurement operators together with their outcome frequencies to reconstruct physical density operators~\cite{ma2025qat}. Masked autoencoder pretraining further learns representations for state reconstruction from incomplete and noisy frequency data~\cite{ma2026ilr}.

Learned denoisers have also been combined with standard reconstruction methods, either to refine their output or to act as priors within their iterations. Attention-based networks refine estimates obtained by linear inversion or MLE from informationally complete measurements~\cite{palmieri2024attention}. Hybrid-denoising QST (\mbox{HD-QST}) alternates measurement-residual updates with a combination of a neural denoiser pretrained on natural images and a low-rank denoiser, weighted by reinforcement learning, and was tested with incomplete Pauli measurements on several quantum processors~\cite{xue2025}.

Incomplete measurement data do not generally identify a unique density operator. Prior information can reduce the measurements needed to identify a state~\cite{heinosaari2013} and enters Bayesian estimators through the prior distribution~\cite{blumekohout2010,granade2016}. Diffusion models learn a distribution by training a network to remove Gaussian noise of varying strength from its samples and generate new samples by repeated denoising~\cite{sohldickstein2015,ho2020}; for inverse problems, their denoisers can supply such prior information, either conditioned directly on the data~\cite{saharia2022sr3} or guided during reconstruction by a measurement likelihood~\cite{chung2023dps}. Related quantum diffusion models generate pure- and mixed-state ensembles~\cite{zhang2024,kwun2025}. Classical diffusion models have also learned the measurement-outcome distribution of a fixed state from finite samples~\cite{wang2023} and generated measurement records conditioned on Hamiltonian parameters to estimate ground-state properties~\cite{tang2025}.

Here, diffusion is used not to generate states or measurement records but to reconstruct states: unlike the direct maps and masked-frequency pretraining above~\cite{ma2025qat,ma2026ilr}, the network learns to remove Gaussian noise of varying strength from a target-state representation, conditioned on the measurement data. Our estimator, \mbox{Diff-QST}, learns from an ensemble of input states to map pairs $(P_j,y_j)$ of Pauli observables and their empirical expectation values (Fig.~\ref{fig:forward}) to a physical density operator; one network per system size and input-state ensemble handles different observable subsets and shot counts. Reconstruction starts from a Gaussian sample rather than an initial tomographic estimate~\cite{palmieri2024attention} and uses a fixed sequence of two or three denoising steps without a convergence test. Unlike HD-QST, it uses learned conditioning instead of explicit measurement-residual updates. We investigate whether this approach, with its short reconstruction process, can improve fidelity over MLE and SDP while reducing reconstruction time after training.

Compared with MLE and SDP, which use no information learned from an input-state ensemble, \mbox{Diff-QST} achieves higher mean fidelity in all $120$ nominal conditions tested, $96$ of which use incomplete Pauli sets. These conditions span two to five qubits, pure and mixed input-state ensembles, $20\%$--$100\%$ of the nonidentity Pauli observables, and $10^2$--$10^4$ shots per observable, at fixed nominal preparation-channel strengths and readout-error rates. At batch size $1$, \mbox{Diff-QST} on GPU is faster than both references in condition-averaged time at every system size; at four and five qubits, \mbox{Diff-QST} is faster in every condition, even on the same CPU resources. Batching further reduces the GPU time per state.

As a supporting comparison, an architecture-matched regression baseline, a single-pass deterministic regressor trained without diffusion~\cite{saharia2022sr3}, tests the benefit of diffusion beyond direct learned reconstruction. \mbox{Diff-QST} has higher condition-averaged fidelity at four and five qubits; the regression baseline has higher fidelity at two and three qubits and reconstructs faster at every tested size. Section~\ref{sec_result_controls} and Appendix~\ref{app_dr} examine the roles of repeated denoising and measurement budgets in this comparison.
Section~\ref{sec_result_shift} repeats both comparisons under changes in the state distribution, readout calibration, and preparation noise, testing how far the nominal results extend.

\section{State reconstruction from Pauli data}\label{sec_method}

\begin{figure*}[!t]
  \centering
  \includegraphics[width=\textwidth]{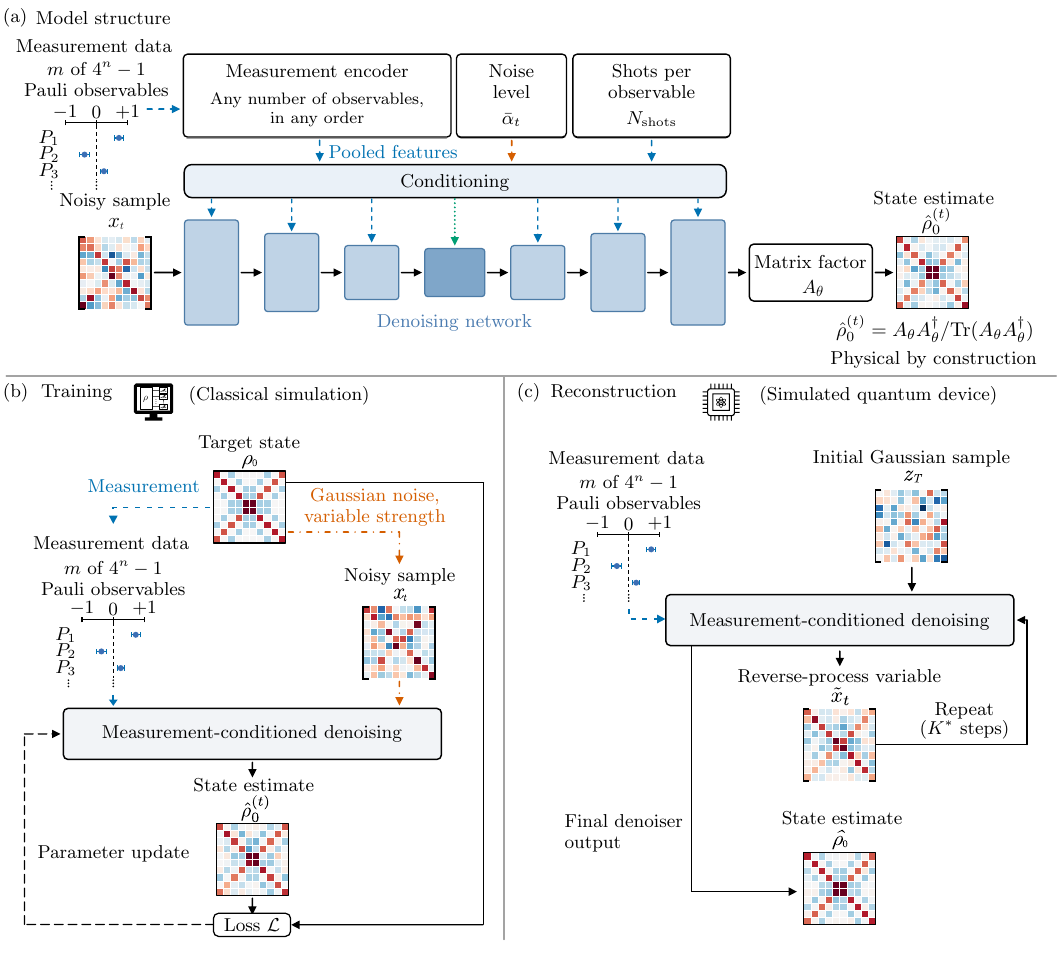}
  \caption{Measurement-conditioned diffusion for quantum state reconstruction. (a) The denoising map takes observable-value pairs, the noise level $\bar\alpha_t$, and the shot count. Blue dashed arrows denote measurement or shot-count paths, orange dash-dotted arrows denote the noise level or Gaussian noise, and the green dotted arrow denotes the measurement features entering at the coarsest resolution. The output is a physical density operator, Eq.~(\ref{eq:factor}). (b) Training minimizes the loss $\mathcal L$ between the state estimate $\hat\rho_0^{(t)}$ and the target state $\rho_0$, given the measurement data and the noisy sample $x_t$. (c) Reconstruction starts from an initial Gaussian sample $z_T$ and uses the same measurements at each of $K^*$ validation-selected denoising steps. Density operators are displayed through their real parts; $x_t$, $z_T$, and the reverse-process variable $\tilde x_t$ are real arrays, not density operators.}
  \label{fig:model}
\end{figure*}

Our target is the physical state $\rho_0=\mathcal{E}(\rho_\mathrm{in})$ after preparation noise and before readout (Fig.~\ref{fig:forward}). Accuracy is quantified by the squared quantum state fidelity $F(\rho,\rho')=\big(\operatorname{Tr}\sqrt{\sqrt{\rho}\,\rho'\sqrt{\rho}}\big)^2$~\cite{jozsa1994}. All comparisons use simulated data and the known target $\rho_0$.

\subsection{Input-state ensembles and measurement model}\label{sec_method_data}

For $n=2,\ldots,5$ qubits with Hilbert-space dimension $d=2^n$, the pure input states are Haar-random and the mixed input states are depolarized pure states~\cite{byrd2008}, with $d-1$ equal eigenvalues. For the mixed input states, the normalized purity $u=[\operatorname{Tr}(\rho_\mathrm{in}^2)-1/d]/(1-1/d)$ is uniform on $[0.55,1]$. Local depolarizing, amplitude-damping, and phase-damping channels, with nominal per-qubit depolarizing probability $p_\mathrm{dep}=0.030$ and damping strengths $\gamma=\lambda=0.040$, act on $\rho_\mathrm{in}$ to produce the targets $\rho_0$. The labels \emph{pure} and \emph{mixed} refer to $\rho_\mathrm{in}$; both target ensembles are mixed. Channel strengths vary by up to $\pm10\%$ during training and take their nominal values for validation and the main comparison. Over the training states, preparation noise lowers the median target purity for pure input states to $0.828$, $0.730$, $0.642$, and $0.564$ at $n=2,3,4,5$, respectively. For each system size and input-state ensemble, $100{,}000$ training, $10{,}000$ validation, and $300$ test states are drawn independently; Appendix~\ref{app_data} specifies the distributions and noise channels and gives the target-purity ranges of the training states.

The target state admits the Pauli expansion
\begin{equation}
  \rho_0=\frac{1}{d}\left(I+\sum_{j=1}^{4^n-1}\langle P_j\rangle P_j\right),
  \qquad \langle P_j\rangle=\operatorname{Tr}(\rho_0P_j),
  \label{eq:pauli_expansion}
\end{equation}
where $P_j$ runs over the $4^n-1$ nonidentity elements of $\{I,X,Y,Z\}^{\otimes n}$, which form the complete Pauli set.
Their exact expectation values determine an arbitrary state. We measure $m=\lfloor r(4^n-1)\rfloor$ of these observables, drawn uniformly at random without replacement, where $r\in(0,100\%]$ is the observable sampling ratio. For an incomplete Pauli set ($r<100\%$), the expectation values are generally consistent with multiple physical states.

Readout is modeled by independent asymmetric bit flips~\cite{bravyi2021}: at the nominal rates, a true $1$ is observed as $0$ with probability $\varepsilon_{01}=0.045$, and a true $0$ as $1$ with probability $\varepsilon_{10}=0.030$. This replaces $P_j$ with the effective observable $\tilde P_j$, obtained by substituting $\eta P_j^{(q)}+\delta I$ for each nonidentity single-qubit factor $P_j^{(q)}$ of $P_j$, where $\eta=1-\varepsilon_{01}-\varepsilon_{10}$ and $\delta=\varepsilon_{01}-\varepsilon_{10}$ (Appendix~\ref{app_data}). For each observable, the observed parity is the product of the local $\pm1$ outcomes at its nonidentity factors. We sample it $N_\mathrm{shots}$ times; with $n_j^+$ the number of $+1$ outcomes,
\begin{equation}
 \begin{aligned}
 n_j^+&\sim\operatorname{Binomial}\!\left(N_\mathrm{shots},
 \frac{1+\operatorname{Tr}(\rho_0\tilde P_j)}{2}\right),
 \\
 y_j&=\frac{2n_j^+}{N_\mathrm{shots}}-1.
 \end{aligned}
 \label{eq:measurement_counts}
\end{equation}
The reconstruction input consists of the measurement data $\mathcal M=\{(P_j,y_j)\}_{j=1}^m$ and the shot count $N_\mathrm{shots}$. Each observable-value pair contains a Pauli observable $P_j$, rather than its effective counterpart $\tilde P_j$, and its empirical expectation value $y_j$.
Instead of the full distribution of multi-qubit outcomes, we retain only the mean parity $y_j$, which contains both readout error and shot noise. Each observable is thus acquired separately, as in compressed-sensing tomography with randomly chosen Pauli observables~\cite{gross2010}; measurements in local Pauli bases would instead yield the parities of several observables from each shot.

\subsection{Conditional diffusion estimator}\label{sec_method_model}

The reconstruction map, shown schematically in Fig.~\ref{fig:model}, is a denoising diffusion probabilistic model (DDPM)~\cite{sohldickstein2015,ho2020} conditioned on the measurement data. In the forward process of a DDPM, Gaussian noise of controlled strength is added to a target representation, and the network is trained to remove it. This supplies denoising tasks over a range of noise strengths, distinct from the shot noise and readout errors in the conditioning data. Reconstruction, the reverse process, applies the learned map repeatedly from an initial Gaussian sample.
The network operates on a real representation, in which the target becomes $x_0=(\operatorname{Re}\rho_0,\operatorname{Im}\rho_0)$, an array of two real $d\times d$ channels. The forward process produces the noisy sample
\begin{equation}
  x_t=\sqrt{\bar{\alpha}_t}\,x_0+\sqrt{1-\bar{\alpha}_t}\,\epsilon,
  \label{eq:forward}
\end{equation}
where $\epsilon\sim\mathcal{N}(0,I)$ is standard Gaussian noise, $t=1,\dots,T$ indexes the $T$ diffusion steps, $\bar{\alpha}_t=\prod_{i=1}^{t}(1-\beta_{i})$, called the noise level as in Ref.~\cite{saharia2022sr3}, decreases from near unity to near zero as more noise is added, following the cosine schedule of Ref.~\cite{nichol2021}, and $\beta_t$ is the variance of the noise added at step $t$. The array $x_t$ is not a density operator, since Eq.~(\ref{eq:forward}) perturbs the real and imaginary parts independently and does not preserve Hermiticity, positivity, or unit trace.

The denoiser $f_\theta$, with trainable parameters $\theta$, estimates the target representation from $x_t$, $\bar\alpha_t$, $\mathcal M$, and $N_\mathrm{shots}$. The observable-value pairs $(P_j,y_j)$ retain which physical quantities were measured, and the conditioning is invariant to their ordering. Figure~\ref{fig:model}(a) shows the U-Net denoiser~\cite{ronneberger2015}, which processes multiple resolutions. A measurement encoder maps any number of observable-value pairs to measurement features, which modulate the U-Net and enter its coarsest resolution, the bottleneck, through attention. Appendix~\ref{app_model} specifies layer dimensions and training settings.

The output layer returns two channels, taken as the real and imaginary parts of a $d\times d$ matrix $A_\theta$, from which the state estimate at diffusion step $t$ is formed as
\begin{equation}
  \hat{\rho}_0^{(t)} =\frac{A_\theta A_\theta^{\dagger}}{\operatorname{Tr}(A_\theta A_\theta^{\dagger})}. \label{eq:factor}
\end{equation}
For $A_\theta\neq0$, Eq.~(\ref{eq:factor}) yields a Hermitian, positive-semidefinite, unit-trace operator~\cite{james2001}. Every state estimate is therefore physical without a penalty or projection step. Physicality alone does not make reconstruction from an incomplete Pauli set unique; the estimator also relies on structure learned from the input-state ensemble.

The training objective is
\begin{equation}
  \begin{aligned}
  &\mathcal{L}(\theta)=\mathbb{E}\!\biggl[\frac{1}{2d^{2}}\Bigl(\\
  &\quad\big\lVert f_\theta(x_t,\bar{\alpha}_t,\mathcal{M},N_\mathrm{shots})-x_0\big\rVert_{F}^{2}\\
  &\quad+\mu\bigl[1-F(\hat\rho_0^{(t)},\rho_0)\bigr]
  \Bigr)\biggr],
  \end{aligned}
  \label{eq:loss}
\end{equation}
which combines the mean squared Hilbert-Schmidt (Frobenius) distance to the target with the infidelity weighted by $\mu$. Here $f_\theta$ returns the real representation of $\hat\rho_0^{(t)}$ in Eq.~(\ref{eq:factor}), and the expectation is taken over the training states, uniformly sampled diffusion steps $t$, the Gaussian noise $\epsilon$, and the measurement data. The training targets are fixed, whereas the diffusion noise and the measurement data are redrawn throughout training. During training, the observable sampling ratio varies between $20\%$ and $100\%$, and the shot count is drawn uniformly from $\{10^2,10^3,10^4,\infty\}$, where $\infty$ denotes the infinite-shot limit $y_j=\operatorname{Tr}(\rho_0\tilde P_j)$, which retains readout error but has no shot noise. All nominal test conditions therefore lie within the training range (Appendix~\ref{app_training}).

Reconstruction runs the reverse process from an initial Gaussian sample $\tilde x_T=z_T\sim\mathcal N(0,I)$ through $K$ denoising steps spanning the full diffusion range [Fig.~\ref{fig:model}(c)]. Between scheduled indices $t'<t$, $\tilde x_{t'}$ is sampled from the forward process's Gaussian conditional distribution given $\tilde x_t$ and $x_0$, with $x_0$ replaced by the predicted state's real representation. This strided posterior update is ancestral sampling~\cite{ho2020} restricted to $K$ of the $T$ diffusion steps~\cite{nichol2021}, and the final denoiser output gives $\hat\rho_0$. The number of denoising steps $K^*$ is selected on validation data from $K=2,\ldots,8$, subject to the constraint that it does not decrease with $n$ within each input-state ensemble (Appendix~\ref{app_kstar}). The selected values, $K^*=2$ at $n=2$ and $K^*=3$ at $n=3,4,5$, are used for all measurement conditions and distribution-shift tests. Excluding $K=1$, which returns only the denoiser prediction from the initial Gaussian sample, guarantees at least one reverse-process update, and $K=1$ is evaluated separately as a control (Sec.~\ref{sec_result_controls}). Appendix~\ref{app_sampler} compares this sampler with a deterministic one and with other initial samples and injected-noise scales; none of these is better in every condition for all eight networks or raises the condition-averaged fidelity by more than $0.0053$, and some lower it by more than $0.05$, so ancestral sampling is retained.

\subsection{Maximum-likelihood and least-squares estimation}\label{sec_method_reference}

The reference estimators, MLE and SDP, use no learned information about the input-state ensemble. They receive the same data as \mbox{Diff-QST} and use the nominal readout model through the effective observables $\tilde{P}_j$, as in readout-aware tomography with calibrated measurement effects~\cite{aasen2024}; \mbox{Diff-QST} instead incorporates nominal readout error through training.
MLE maximizes the likelihood with the diluted $R\rho R$ iteration~\cite{rehacek2007}, treating each observable as a two-outcome measurement with effects $E_{j}^{\pm}=(I\pm\tilde{P}_j)/2$ and counts recovered from $y_j$.

The SDP estimator solves
\begin{equation}
  \min_{\rho\succeq0,\ \operatorname{Tr}\rho=1}\ \sum_{j=1}^{m}
  w_j\left[\operatorname{Tr}(\rho\tilde{P}_j)-y_j\right]^{2}
  +\tau\left\lVert\rho-\frac{I}{d}\right\rVert_{F}^{2},
  \label{eq:sdp}
\end{equation}
where $\|\cdot\|_F$ is the Frobenius norm. The weights $w_j$ are inverse estimates of the sampling variance of $y_j$, normalized to unit mean, with the outcome probability taken as the posterior mean under the Jeffreys prior $\operatorname{Beta}(1/2,1/2)$~\cite{jeffreys1946}, so that each weight is finite even when all outcomes of an observable agree. For unit-trace $\rho$, the regularization term in Eq.~(\ref{eq:sdp}) is the purity penalty $\tau[\operatorname{Tr}(\rho^2)-1/d]$~\cite{mu2020}, with $\tau=10^{-6}$. It makes the objective strictly convex, so the estimate is unique even for incomplete Pauli sets, but it can also alter the measurement fit. Appendix~\ref{app_compute} gives the weights, solver settings, and numerical checks.

\subsection{Architecture-matched regression baseline}\label{sec_method_dr}

To assess the benefit of diffusion beyond direct learned reconstruction, we train an architecture-matched regression baseline for each system size and input-state ensemble, as in evaluations of conditional diffusion models~\cite{saharia2022sr3}. This single-pass deterministic regressor retains the corresponding denoiser architecture, measurement encoder, physical output parametrization of Eq.~(\ref{eq:factor}), and composite training loss, but makes a single prediction with its noisy-sample input set to zero and its noise level set to one, the value at which no Gaussian noise is added; it is trained without Gaussian perturbation at randomly sampled diffusion steps and reconstructs without a reverse process. To make the use of diffusion the only difference between the two estimators, each regression network is trained with the data, optimizer, epoch limit, and early-stopping rules of the corresponding \mbox{Diff-QST} network and with its learning rate, ratio curriculum, and infidelity weight $\mu$, which were chosen on validation data for \mbox{Diff-QST}. The weights of the two estimators are selected independently on validation data. The regression baseline uses one network evaluation, whereas \mbox{Diff-QST} uses one for each of its two or three denoising steps.

\section{Results}\label{sec_result}
\raggedbottom

\begin{figure*}[t]
  \centering
  \includegraphics[width=\textwidth]{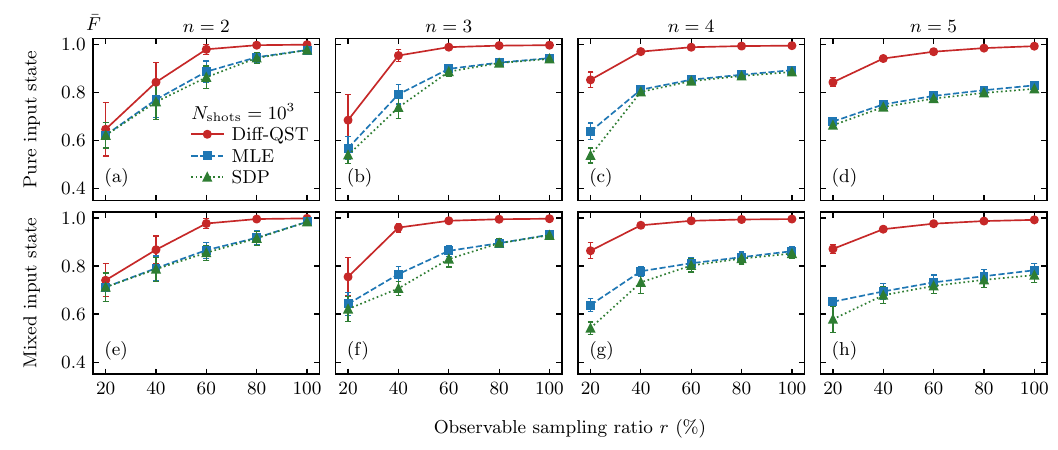}
  \caption{Mean fidelity of \mbox{Diff-QST}, MLE, and SDP. The top row, (a)--(d), corresponds to pure input states, the bottom row, (e)--(h), to mixed input states, and the columns to $n=2$, $3$, $4$, and $5$. All results use $10^3$ shots per observable with nominal preparation-channel strengths and readout-error rates, where $r$ denotes the measured fraction of the $4^n-1$ nonidentity Pauli observables. Markers give $\bar F$ over $300$ test states (Sec.~\ref{sec_result}), and vertical bars mark $\pm1$ standard deviation across these states, truncated at $0$ and $1$.}
  \label{fig:accuracy}
\end{figure*}

For each system size $n=2,\ldots,5$ and each input-state ensemble, the estimators are compared at three shot counts, $N_\mathrm{shots}=10^2$, $10^3$, and $10^4$, and at five observable sampling ratios from $r=20\%$ to $100\%$ in steps of $20\%$. This gives $120$ nominal conditions, of which the $96$ with $r<100\%$ use incomplete Pauli sets. In each condition, all estimators reconstruct the same $300$ test states (Sec.~\ref{sec_method_data}) from the same measurement outcomes. For $r<100\%$, each state's fidelity is averaged over three observable subsets shared by all test states of the condition, and the mean fidelity $\bar F$ averages these per-state values over the test states. Uncertainties of fidelity differences are $95\%$ bootstrap confidence intervals (CIs) over test states~\cite{efron1979}, conditional on the trained networks and observable subsets; ``both CIs'' refers to those relative to MLE and to SDP (Appendix~\ref{app_stats}).

\subsection{Fidelity relative to MLE and SDP}\label{sec_result_accuracy}

In all $120$ nominal conditions, \mbox{Diff-QST} has higher mean fidelity than both MLE and SDP, with both CIs above zero (Table~\ref{tab:nominal_summary}). After Holm's correction for the $240$ comparisons~\cite{holm1979}, all differences remain significant at family-wise level $0.05$ (Appendix~\ref{app_stats}). Figure~\ref{fig:accuracy} shows this comparison at $10^3$ shots per observable. This ordering is not limited to mean fidelities: at four and five qubits, \mbox{Diff-QST} has higher fidelity than both MLE and SDP for every test state in $59$ of the $60$ conditions and for $299$ of the $300$ states in the remaining one ($n=5$, pure input states, $10^2$ shots, $r=20\%$). At two and three qubits, the fraction of such states is at least $0.62$ and $0.83$, respectively, with the lowest values at $r=20\%$.

\begin{table}[b]
  \caption{Fidelity difference $\Delta\bar F$ of \mbox{Diff-QST} from the better of MLE and SDP over the full nominal grid. Entries give the median and, in brackets, the minimum over the $15$ shot-ratio conditions of each system size and input-state ensemble. Both CIs lie above zero in all $120$ conditions.}
  \label{tab:nominal_summary}
  \begin{ruledtabular}
  \begin{tabular}{ccc}
  $n$ & Pure input states & Mixed input states \\
  \colrule
  $2$ & $0.0603\;[0.0019]$ & $0.0770\;[0.0007]$ \\
  $3$ & $0.1161\;[0.0119]$ & $0.1243\;[0.0074]$ \\
  $4$ & $0.1585\;[0.0321]$ & $0.1912\;[0.0386]$ \\
  $5$ & $0.1752\;[0.0631]$ & $0.2294\;[0.0894]$ \\
  \end{tabular}
  \end{ruledtabular}
\end{table}

Table~\ref{tab:nominal_summary} summarizes $\Delta\bar F=\bar F^{(\text{Diff-QST})}-\max\{\bar F^{(\mathrm{MLE})},\bar F^{(\mathrm{SDP})}\}$ across all $120$ nominal conditions. Equivalently, $\Delta\bar F$ is the smaller of the differences from MLE and from SDP in each condition. For the $96$ incomplete-Pauli conditions, $\Delta\bar F\ge0.021$, with medians of $0.120$ for pure and $0.158$ for mixed input states. For each input-state ensemble, the median and minimum of $\Delta\bar F$ over the $15$ conditions of a system size increase with $n$, although individual conditions follow no single monotonic trend in $n$. Two other quantities change with $n$: the target purity decreases (Sec.~\ref{sec_method_data}), and at fixed $r$ and $N_\mathrm{shots}$ the total number of shots, $mN_\mathrm{shots}$, increases. The trend in $n$ therefore does not isolate a system-size effect at fixed target purity or total measurement budget.

At fixed $r$, the amount of data per state parameter also grows with $n$. The number of measured observables per real parameter of the input-state family, $m/d_\psi$, exceeds $3$ throughout the five-qubit grid. Appendix~\ref{app_coverage} therefore analyzes $\Delta\bar F$ against $m/d_\psi$, with four- and five-qubit networks separately retrained to reach ratios below the $20\%$ training bound. For these retrained networks at $N_\mathrm{shots}\ge10^3$, the largest observed fidelity difference occurs at $m/d_\psi=1.47$--$2.02$ in seven of eight combinations of system size, input-state ensemble, and shot count, corresponding to measurement counts approximately between the dimensional lower bound $d_\psi$ for identifying every member of the family and the roughly $2d_\psi$ that topological obstructions can require~\cite{heinosaari2013}.

\begin{figure*}[t]
  \centering
  \includegraphics[width=7.05in]{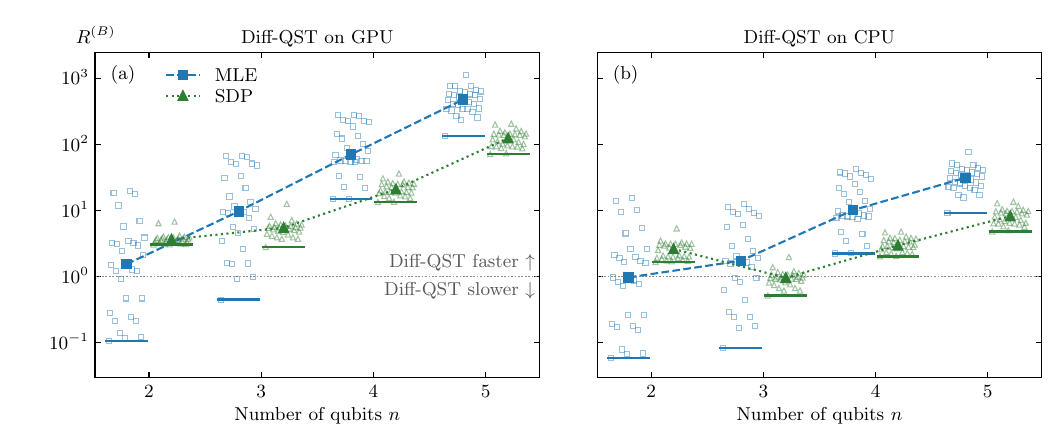}
  \caption{Reconstruction time of MLE and SDP relative to \mbox{Diff-QST} at batch size $1$. For reference $B$, the ratio $R^{(B)}$ divides its mean time over the same $15$ states and observable subsets by \mbox{Diff-QST}'s mean time per state. \mbox{Diff-QST} runs on (a) the GPU and (b) the CPU, and MLE and SDP use the $32$-core CPU configuration in both panels. Open symbols, offset horizontally for visibility, are the $30$ nominal conditions per system size. Filled symbols and connecting lines show their medians, short horizontal bars mark the minima, and the line at unity indicates equal times.}
  \label{fig:time}
\end{figure*}

With the complete Pauli set, the fidelity difference narrows as the number of shots per observable increases. From $10^2$ to $10^4$ shots, the mean fidelity rises by $0.068$--$0.287$ for MLE and $0.069$--$0.284$ for SDP, but by only $0.010$--$0.043$ for \mbox{Diff-QST}, so $\Delta\bar F$ decreases from $0.060$--$0.333$ to $0.0007$--$0.089$, and every minimum in Table~\ref{tab:nominal_summary} occurs at $10^4$ shots with the complete Pauli set. Three shot counts do not separate statistical error from reconstruction bias, nor do they show whether the fidelity of \mbox{Diff-QST} saturates below unity at larger shot counts.

At $n=2$ and $r=20\%$, the advantage is not robust to the choice of observables. With only three observables measured, averaging over all $455$ possible subsets leaves a difference of $0.004$--$0.010$ for pure input states, and both CIs include zero at $10^3$ and $10^4$ shots. Some individual subsets reverse the ordering (Appendix~\ref{app_stats}).

A sensitivity check with hedged MLE, which multiplies the likelihood by a power of $\det\rho$ to keep the estimate full rank~\cite{blumekohout2010hedged}, narrows the fidelity differences, with \mbox{Diff-QST} retaining higher mean fidelity in $91$ of the $96$ incomplete-Pauli and all $24$ complete-Pauli conditions (Appendix~\ref{app_hmle}). With the trace distance $D_{\rm tr}(\rho,\rho')=\tfrac12\lVert\rho-\rho'\rVert_1$~\cite{nielsen2010} instead of fidelity, \mbox{Diff-QST} is closer to the target than both references, with both CIs below zero, in $112$ of the $120$ nominal conditions; the references are closer, with both CIs above zero, in four (Appendix~\ref{app_distance}): three with the complete Pauli set at $10^4$ shots and $n\le3$, and one at $n=5$, $10^4$ shots, and $r=40\%$, where the rank-deficient MLE and SDP estimates are penalized more by fidelity than by trace distance.

\subsection{Reconstruction time relative to MLE and SDP}\label{sec_result_cost}

For a fixed network, number of measured observables $m$, and $K^*$, \mbox{Diff-QST} applies the same operations to every state, whereas MLE and SDP iterate to state-dependent convergence: for five-qubit pure input states at $10^3$ shots and $r=60\%$, MLE takes $1.83$--$12.0$~s per state and SDP $0.657$--$0.776$~s. We time the estimators on the same $15$ test states and observable subsets at batch size $1$, excluding training and measurement acquisition (Appendix~\ref{app_timing}). \mbox{Diff-QST} uses one NVIDIA H200 NVL GPU or the same CPU resource configuration as the references: two AMD EPYC 9135 processors.

When condition-level mean times are averaged equally over the $30$ nominal conditions at each system size, GPU \mbox{Diff-QST} is faster than both references for $n=2,\ldots,5$. Figure~\ref{fig:time} shows the distribution of the condition-level ratios. At five qubits, the median ratios relative to MLE and SDP are $482$ and $125$, with minima of $135$ and $71.9$. At four and five qubits and $10^3$ shots, GPU \mbox{Diff-QST} takes $4.70$--$5.08$~ms per state for incomplete Pauli sets. The speed advantage holds in every four- and five-qubit condition, including when \mbox{Diff-QST} uses the same CPU resources [Fig.~\ref{fig:time}(b)]. At two and three qubits, GPU \mbox{Diff-QST} is faster than SDP throughout, but slower than MLE in $11$ and $2$ conditions, respectively [Fig.~\ref{fig:time}(a)]. These ratios describe the tested implementations and hardware. Accelerated projected-gradient algorithms reach the maximum-likelihood estimate faster than the diluted iteration used here~\cite{shang2017,bolduc2017}, and a faster solver could likewise shorten the SDP times; such algorithms would not change the converged estimates wherever these are unique (Appendix~\ref{app_timing}).

\begin{table*}[t]
  \caption{Comparison with the regression baseline (Reg) over the $12$ finite-shot conditions with incomplete Pauli sets. Entries are condition-averaged mean fidelities; $\bar F^{(K)}$ denotes \mbox{Diff-QST} with $K$ denoising steps. \mbox{Diff-QST} uses $K^*$, $2$ at $n=2$ and $3$ otherwise; the $K=1$ control evaluates the same trained networks with one network evaluation at $n=3,4,5$. Differences are computed from unrounded means and given with $95\%$ CIs over test states. The CIs quantify test-state uncertainty for the reported trained networks; sensitivity to training initialization is assessed separately using three additional training runs per estimator and input-state ensemble at each system size (Appendix~\ref{app_dr}).}
  \label{tab:dr_macro}
  \begin{ruledtabular}
  \begin{tabular}{crrrlll}
  $n$ & Regression & $K=1$ & $K^*$ & $\bar F^{(K^*)}-\bar F^{(\mathrm{Reg})}$ [95\% CI] & $\bar F^{(1)}-\bar F^{(\mathrm{Reg})}$ [95\% CI] & $\bar F^{(K^*)}-\bar F^{(1)}$ [95\% CI] \\
  \colrule
  \multicolumn{7}{l}{Pure input states} \\
  2 & $0.8689$ & --- & $0.8612$ & $-0.0077\,[-0.0111,-0.0044]$ & --- & --- \\
  3 & $0.9007$ & $0.8693$ & $0.8925$ & $-0.0081\,[-0.0102,-0.0060]$ & $-0.0313\,[-0.0327,-0.0300]$ & $+0.0232\,[+0.0210,+0.0254]$ \\
  4 & $0.9041$ & $0.9296$ & $0.9303$ & $+0.0262\,[+0.0251,+0.0273]$ & $+0.0255\,[+0.0245,+0.0266]$ & $+0.0007\,[+0.0001,+0.0012]$ \\
  5 & $0.8405$ & $0.9142$ & $0.9080$ & $+0.0675\,[+0.0642,+0.0708]$ & $+0.0737\,[+0.0704,+0.0770]$ & $-0.0062\,[-0.0066,-0.0057]$ \\
  \colrule
  \multicolumn{7}{l}{Mixed input states} \\
  2 & $0.8952$ & --- & $0.8891$ & $-0.0061\,[-0.0081,-0.0042]$ & --- & --- \\
  3 & $0.9173$ & $0.8935$ & $0.9118$ & $-0.0054\,[-0.0071,-0.0037]$ & $-0.0237\,[-0.0249,-0.0225]$ & $+0.0183\,[+0.0165,+0.0201]$ \\
  4 & $0.9106$ & $0.9353$ & $0.9326$ & $+0.0220\,[+0.0210,+0.0230]$ & $+0.0247\,[+0.0239,+0.0255]$ & $-0.0027\,[-0.0032,-0.0022]$ \\
  5 & $0.8976$ & $0.9217$ & $0.9179$ & $+0.0203\,[+0.0187,+0.0219]$ & $+0.0241\,[+0.0226,+0.0256]$ & $-0.0038\,[-0.0043,-0.0033]$ \\
  \end{tabular}
  \end{ruledtabular}
\end{table*}

In $12$ four- and five-qubit conditions at $10^3$ shots, stopping MLE or SDP early removes neither the fidelity nor the time advantage (Appendix~\ref{app_timing}): even the best tested stopping point, selected with knowledge of the targets, leaves their mean fidelity $0.09$--$0.20$ below that of \mbox{Diff-QST}, and stopping each once its mean fidelity comes within $0.001$ of the converged value leaves them $13$--$676$ times slower than GPU \mbox{Diff-QST}. When many states require reconstruction, GPU batching further reduces \mbox{Diff-QST}'s time per state, by a factor of $9.3$--$87.1$ at batch size $256$ relative to batch size $1$ across the tested networks and observable sampling ratios.

\subsection{Comparison with the regression baseline}\label{sec_result_controls}

Over the $12$ finite-shot conditions with incomplete Pauli sets (Table~\ref{tab:dr_macro}), \mbox{Diff-QST} has higher condition-averaged fidelity than the regression baseline at four and five qubits for both input-state ensembles, by $0.0203$--$0.0675$, whereas the regression baseline has higher fidelity at two and three qubits; all eight CIs exclude zero. At four and five qubits, the advantage is largest at $r\le40\%$: averaged over shot counts, it is $0.028$--$0.134$ at $r=20\%$ but at most $0.014$ with the complete Pauli set (Appendix~\ref{app_dr}). Retraining both estimators from three further initializations at $n=2,3,4$ reproduces the two- and three-qubit ordering in all $12$ further pairs; at four qubits, three of the four pairs per input-state ensemble, including the reported one, favor \mbox{Diff-QST}. At five qubits, all three further pairs per input-state ensemble favor \mbox{Diff-QST}, with CIs above zero (Appendix~\ref{app_dr}). The regression baseline, however, reconstructs faster. At batch size $1$ on the same GPU, the four- and five-qubit \mbox{Diff-QST} networks take about $2.2$--$2.3$ times as long (Appendix~\ref{app_timing}).

A separate $K=1$ control (Appendix~\ref{app_dr_steps}) evaluates the same trained networks with a single denoising step. It retains the four- and five-qubit fidelity advantage over the regression baseline, so this advantage does not require repeated network evaluation. Because the $K=1$ estimator and the regression baseline still differ in training and in their noisy-sample and noise-level inputs, this comparison does not identify which element of diffusion training produces the advantage. Relative to $K=1$, the selected $K^*=3$ raises the condition-averaged fidelity of both three-qubit networks, by $0.0183$ and $0.0232$. At four and five qubits it changes the fidelity little: a gain of $0.0007$ for the four-qubit network for pure input states and losses of $0.0027$--$0.0062$ for the other three networks. Because selection excludes $K=1$ (Sec.~\ref{sec_method_model}), $K^*$ is not an optimum over all step counts.

At five qubits, we also compare the smallest evaluated total shot budget, $N_\mathrm{tot}=mN_\mathrm{shots}$, at which each estimator attains a specified mean fidelity, over $21$ combinations of $m$ and $N_\mathrm{shots}$ per input-state ensemble, including $N_\mathrm{shots}=200$ and $500$, which lie between the shot counts used in training (Appendix~\ref{app_dr_shots}). For $\bar F\ge0.95$, \mbox{Diff-QST} requires $N_\mathrm{tot}=102{,}300$ for both input-state ensembles, whereas the regression baseline requires $204{,}600$ for pure and $306{,}500$ for mixed input states, so \mbox{Diff-QST} uses $50\%$ and $67\%$ fewer total shots; for $\bar F\ge0.90$, it uses $40\%$ and $25\%$ fewer. A finer grid than the evaluated one could change these minima.

\begin{figure*}[t]
  \centering
  \includegraphics[width=\textwidth]{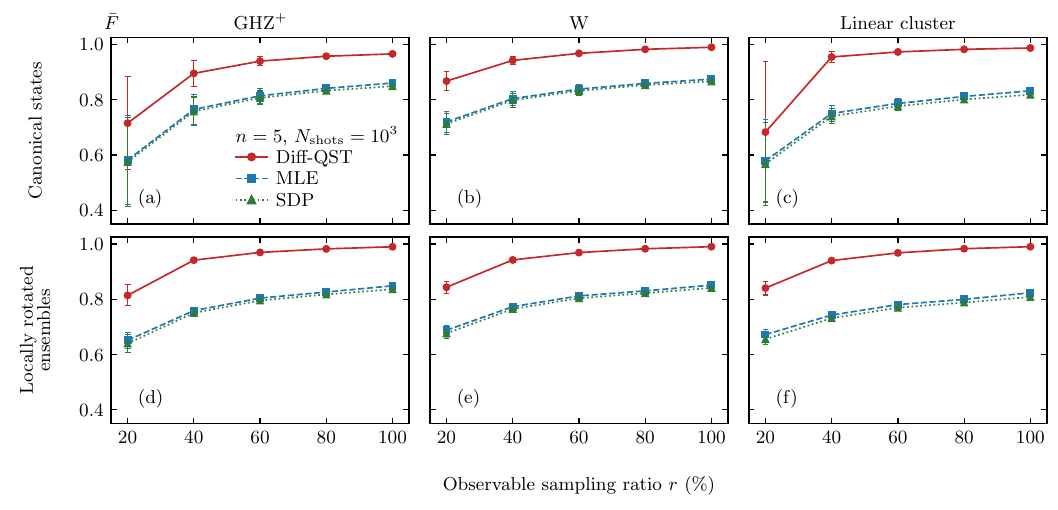}
  \caption{Reconstruction fidelity for five-qubit structured input states at $10^3$ shots per observable, using the network for pure input states: (a)--(c) canonical GHZ$^{+}$, W, and linear cluster states and (d)--(f) their locally rotated ensembles. Markers show mean fidelities. Vertical bars mark $\pm1$ standard deviation across $100$ independent acquisitions for each canonical state or across the subset-averaged fidelities of $300$ states for each rotated ensemble, truncated at $0$ and $1$. Figure~\ref{fig:structured_heatmap} shows fidelity differences over the full shot-ratio grids for $n=2,\ldots,5$.}
  \label{fig:structured}
\end{figure*}

\subsection{Changes in the state and noise distributions}\label{sec_result_shift}

The fidelity advantage over MLE and SDP is tested next under changes in the state or noise distribution. The networks are applied without retraining and with $K^*$ unchanged, and MLE and SDP retain the nominal readout model. In each test, all estimators use the same targets, observable subsets, and measurement outcomes. Table~\ref{tab:shift_summary} summarizes the results.

\begin{table}[t]
  \caption{Summary of the tests with shifted state or noise distributions. All counts are summed over $n=2,\ldots,5$. For each test, the last two columns give the number of conditions in which both CIs relative to MLE and SDP lie above or below zero. The spectral and readout tests (top) use the $15$ shot-ratio conditions per system size. The structured-state tests (middle) apply these conditions to each of five canonical representatives or three rotated ensembles. The preparation-noise tests (bottom) use $r=20\%$, $60\%$, and $100\%$ at three shot counts. The grids differ, so the counts do not rank the tests.}
  \label{tab:shift_summary}
  \begin{ruledtabular}
  \begin{tabular}{llrrr}
  Change & Input states & Cond. & ${}>0$ & ${}<0$ \\
  \colrule
  Degeneracy lifted & mixed & $60$ & $59$ & $1$ \\
  Lower purity range & mixed & $60$ & $58$ & $2$ \\
  Readout $\times0.5$ & pure, mixed & $120$ & $120$ & $0$ \\
  Readout $\times2$ & pure, mixed & $120$ & $120$ & $0$ \\
  \colrule
  Canonical structured & pure & $300$ & $233$ & $30$ \\
  Rotated structured & pure & $180$ & $171$ & $9$ \\
  \colrule
  Preparation $\times0.25$ & pure & $36$ & $26$ & $10$ \\
  & mixed & $36$ & $35$ & $1$ \\
  Preparation $\times0.5$ & pure & $36$ & $35$ & $1$ \\
  & mixed & $36$ & $36$ & $0$ \\
  \end{tabular}
  \end{ruledtabular}
\end{table}

With nominal noise, we test two shifts of the distribution of mixed input states: (i) replacing the degenerate spectra with randomly generated nondegenerate spectra that have the same distribution of normalized purity, and (ii) lowering the purity range to $u\in[0.30,0.55]$ (Appendix~\ref{app_statedist}). For each shift, both CIs lie above zero in all $48$ incomplete-Pauli conditions. The exceptions on the full grids occur for complete Pauli sets at $10^4$ shots and small system sizes, with a most negative difference of $\Delta\bar F=-0.0036$. For incomplete Pauli sets, the fidelity advantage therefore generalizes beyond the degenerate spectral structure used in training.
Device readout error rates drift between calibrations~\cite{proctor2020,dasgupta2020,hirasaki2023}; to test sensitivity to the resulting readout miscalibration, we halve or double the actual readout rates. For each rescaling, without retraining or updating the assumed readout model, \mbox{Diff-QST} has higher mean fidelity than both references in all $120$ conditions, with both CIs above zero (Appendix~\ref{app_ood_summary}).

The networks for pure input states, whose training and tuning include no structured-state examples, are next applied to Greenberger-Horne-Zeilinger (GHZ)~\cite{greenberger1989}, W~\cite{dur2000}, and linear cluster states~\cite{briegel2001} under nominal noise and the $15$ measurement conditions (Appendix~\ref{app_structured}). Each family is tested in two forms: canonical representatives and locally rotated ensembles. The canonical representatives are the GHZ states with relative phases $0$, $\pi$, and $\pi/2$ (GHZ$^{+}$, GHZ$^{-}$, GHZ$^{i}$), the W state, and the linear cluster state. The rotated ensembles are generated by applying independent Haar-random single-qubit rotations to GHZ$^{+}$, W, and the linear cluster state, which keeps the entanglement of each ideal input but randomizes its orientation relative to the Pauli measurement basis. As an unstructured control, a separate ensemble of Haar-random pure input states, drawn from the same distribution as the test states of the main comparison, is evaluated in the same way as the rotated ensembles.

Every tested canonical representative and rotated ensemble at $n=3,4,5$ has a positive median over the $15$ conditions, although some three-qubit canonical conditions have both CIs below zero. At $n=2$, the canonical GHZ representatives have negative medians, and the rotated ensembles have both CIs below zero at $r=20\%$ for every shot count. The Haar-random ensemble also has both CIs below zero at $n=2$, $r=20\%$ with its own three observable subsets, which differ from those of the main comparison, but above zero after averaging over all $455$ three-observable subsets (Appendix~\ref{app_stats}); the rotated ensembles remain below zero after this averaging (Appendix~\ref{app_structured}), so their reversal is not explained by subset choice. Figure~\ref{fig:structured} shows the five-qubit mean fidelities at $10^3$ shots: at $r=20\%$, \mbox{Diff-QST} attains $0.683$--$0.867$ for the canonical states, well below unity even though the difference from MLE and SDP is positive.

Scaling all preparation-channel parameters to one-half or one-quarter of their nominal values, with readout nominal, shifts the target distribution beyond the training range. Both CIs remain above zero in $36$ (mixed) and $35$ (pure) of the $36$ conditions at half strength, and in $35$ and $26$ conditions at quarter strength. The $10$ quarter-strength reversals for pure input states occur at every system size, in $2$, $2$, $2$, and $4$ of the nine conditions at $n=2$, $3$, $4$, and $5$. They include incomplete Pauli sets, with the most negative difference, $\Delta\bar F=-0.1010$, at $n=5$, $r=20\%$, and $10^4$ shots. The nominal advantage therefore does not extend uniformly to weaker preparation noise. The shifted targets are purer than those in training: at $n=5$, the median target purity for pure input states is $0.87$ and $0.75$ at quarter and half strength, compared with $0.56$ nominally and at most $0.64$ over the training states. Because these tests do not isolate purity from other properties of the targets, the reversals cannot be attributed to purity alone (Appendix~\ref{app_channel}).

Applied without retraining to the same shifted data, the regression baseline has lower mean fidelity than \mbox{Diff-QST} at four and five qubits in $238$ of the $252$ spectral, readout, and preparation-noise conditions and, on average, for the structured states other than the canonical GHZ representatives; at two and three qubits, it has the higher mean fidelity in $178$ of the corresponding $252$ conditions (Appendix~\ref{app_dr_shift}). Under weaker preparation noise, it also reverses relative to MLE and SDP more often than \mbox{Diff-QST}: for pure input states, both CIs lie below zero in $12$ and $4$ of $36$ conditions at quarter and half strength, compared with $10$ and $1$ for \mbox{Diff-QST}.

\section{Conclusion}\label{sec_conclusion}

We have introduced \mbox{Diff-QST}, a measurement-conditioned diffusion estimator that learns state information from an input-state ensemble and reconstructs a density operator from noisy, possibly incomplete Pauli measurements in two or three denoising steps. In simulations of two to five qubits, it has higher mean fidelity than MLE and SDP, which use no learned state information, in all $120$ nominal conditions. Because the number of steps is fixed, reconstruction involves no state-dependent convergence iterations. Once trained, \mbox{Diff-QST} reconstructs individual states on a GPU at a batch size of $1$, achieving a lower condition-averaged time than both references at every system size, and is faster than them in all four- and five-qubit conditions, even on the same CPU resources.

Two conditions limit these advantages. The first is the training distribution, from which \mbox{Diff-QST} learns its state information. For incomplete Pauli sets, the advantage over MLE and SDP persists under the tested changes in mixed-state spectra and readout calibration but can reverse under weaker preparation noise and for some two- and three-qubit structured states. The second is the system size, as the advantage of diffusion over learned regression appears only in the larger tested systems. At four and five qubits, \mbox{Diff-QST} has higher condition-averaged fidelity than the regression baseline, an advantage retained in most of the tested distribution shifts and training initializations. At nominal noise, it does not require repeated denoising, as it is retained with a single denoising step. At five qubits, over the tested numbers of observables and shots per observable, \mbox{Diff-QST} also needs $25\%$--$67\%$ fewer total shots to reach specified fidelities. At two and three qubits, the regression baseline has higher fidelity. It is also faster at every tested size.

These results are limited to simulations of two to five qubits with separately trained networks. They do not establish performance on experimental data, whose preparation and readout noise can differ from those assumed in training, or beyond five qubits, because the number of measured Pauli observables at fixed sampling ratio and the density-operator representation both grow as $4^n$. Within this scope, \mbox{Diff-QST} is useful where fast reconstruction is required. Its speed advantage over MLE and SDP grows with the number of qubits over the tested sizes and, through GPU batching, with the number of states to be reconstructed.

\section*{Acknowledgement}
We acknowledge the use of AI tools to refine the English phrasing and grammar of this manuscript. This work was partly supported by the Institute for Information \& communications Technology Planning \& Evaluation (IITP) grant (RS-2025-02292999), the IITP-ITRC (Information Technology Research Center) grant (RS-2026-25519864), and Creation of the Quantum Information Science R\&D Ecosystem (RS-2023-NR068116) through the National Research Foundation of Korea (NRF), funded by the Korean government (MSIT).

\newpage
\twocolumngrid
\setlength{\intextsep}{12pt plus 2pt minus 2pt}
\setlength{\textfloatsep}{12pt plus 2pt minus 2pt}
\raggedbottom
\appendix

\section{Input-state ensembles and noise model}\label{app_data}

For each system size $n=2,\ldots,5$ and each input-state ensemble, we independently draw $100{,}000$ training states, $10{,}000$ validation states, and $300$ test states. The pure input states are Haar-random. The mixed input states are depolarized pure states~\cite{byrd2008},
\begin{equation}
  \rho_\mathrm{in}=p\,|\psi\rangle\langle\psi|+(1-p)\frac{I}{d},
  \label{eq:dps}
\end{equation}
with $|\psi\rangle$ Haar-random, normalized purity $u$ uniform on $[0.55,1]$, and $p=\sqrt u$. Each state in this family has one large eigenvalue and $d-1$ equal eigenvalues, so the family does not cover arbitrary mixed states.

Preparation noise applies local depolarizing, amplitude-damping, and phase-damping channels in that order to every qubit~\cite{nielsen2010}. This is a phenomenological channel applied once after ideal preparation, not a gate-resolved model of a particular device. The depolarizing channel is
\begin{equation}
  \begin{aligned}
  \mathcal{E}_\mathrm{dep}(\rho)&=(1-p_\mathrm{dep})\,\rho\\
  &\quad+\frac{p_\mathrm{dep}}{3}\left(X\rho X+Y\rho Y+Z\rho Z\right).
  \end{aligned}
  \label{eq:depol}
\end{equation}
The damping channels have Kraus operators
\begin{equation}
  \begin{aligned}
  M_0^{(\mathrm{ad})}&=\begin{pmatrix}1&0\\0&\sqrt{1-\gamma}\end{pmatrix},\\
  M_1^{(\mathrm{ad})}&=\begin{pmatrix}0&\sqrt{\gamma}\\0&0\end{pmatrix},
  \end{aligned}
  \label{eq:amp}
\end{equation}
\begin{equation}
  \begin{aligned}
  M_0^{(\mathrm{ph})}&=\begin{pmatrix}1&0\\0&\sqrt{1-\lambda}\end{pmatrix},\\
  M_1^{(\mathrm{ph})}&=\begin{pmatrix}0&0\\0&\sqrt{\lambda}\end{pmatrix}.
  \end{aligned}
  \label{eq:phase}
\end{equation}
The nominal channel parameters are $p_\mathrm{dep}=0.030$ and $\gamma=\lambda=0.040$; during training, each strength is sampled independently per state within $\pm10\%$ of its nominal value.
Over the training states, which include this variation, the target purity $\operatorname{Tr}(\rho_0^2)$ ranges over $0.763$--$0.926$, $0.664$--$0.860$, $0.584$--$0.751$, and $0.510$--$0.636$ at $n=2,3,4,5$ for pure input states and over $0.532$--$0.912$, $0.428$--$0.833$, $0.354$--$0.741$, and $0.300$--$0.620$ for mixed input states. At $n=5$, the purity of mixed input states has median $0.783$ before the channel and $0.445$ afterward.

Independent asymmetric readout errors replace each Pauli string by
\begin{equation}
  \begin{aligned}
  \tilde{P}_j&=\bigotimes_{q=1}^{n}\tilde{P}_j^{(q)},\\
  \tilde{P}_j^{(q)}&=
  \begin{cases}
    I, & P_j^{(q)}=I,\\[2pt]
    \eta\,P_j^{(q)}+\delta\,I, & \text{otherwise},
  \end{cases}
  \end{aligned}
  \label{eq:readout}
\end{equation}
where $P_j^{(q)}$ is the factor of $P_j$ acting on qubit $q$, $\eta=1-\varepsilon_{01}-\varepsilon_{10}$, and $\delta=\varepsilon_{01}-\varepsilon_{10}$. Finite-shot data follow Eq.~(\ref{eq:measurement_counts}).

\begin{table*}[t]
  \caption{Network dimensions and training settings. U-Net channel counts run from the input resolution to the bottleneck, and Parameters is the number of trainable parameters. Encoder columns specify token dimension, maximum latent-vector count, attention heads, and Transformer layers. $T$ is the number of diffusion steps, $\mu$ is the infidelity weight in Eq.~(\ref{eq:loss}), the learning rate is the initial AdamW value, curriculum is the number of epochs over which the lower bound of the training ratio decreases to $20\%$, and $K^{*}$ is the number of denoising steps selected on validation data under the nondecreasing constraint described in Appendix~\ref{app_kstar}.}
  \label{tab:arch}
  \begin{ruledtabular}
  \begin{tabular}{ccrlrrrrr@{\hspace{14pt}}crrr}
   & & & & & \multicolumn{4}{c}{Measurement encoder} & & & & \\
  \cmidrule{6-9}
   $n$ & Input state & $T$ & U-Net channels & Parameters & Dim. & Latents & Heads & Layers & $\mu$ & Learning rate ($10^{-4}$) & Curriculum & $K^{*}$ \\
  \colrule
   2 & pure  &  60 & 64, 128        & $1{,}656{,}755$  & 128 &  32 & 4 & 1 & 1 & 3.65  &  20 & 2 \\
   3 & pure  &  90 & 128, 256, 512  & $23{,}312{,}659$ & 256 &  64 & 8 & 2 & 4 & 1.48  &  80 & 3 \\
   4 & pure  & 120 & 128, 256, 512  & $30{,}771{,}731$ & 512 &  96 & 8 & 3 & 4 & 0.732 & 120 & 3 \\
   5 & pure  & 180 & 192, 384, 768  & $59{,}006{,}163$ & 512 & 256 & 4 & 4 & 4 & 0.567 & 280 & 3 \\
  \colrule
   2 & mixed &  50 & 128, 256       & $5{,}264{,}563$  & 128 &  64 & 4 & 1 & 2 & 3.49  &  40 & 2 \\
   3 & mixed &  70 & 128, 256, 512  & $31{,}139{,}027$ & 512 &  32 & 8 & 2 & 2 & 1.20  & 160 & 3 \\
   4 & mixed & 130 & 128, 256, 512  & $31{,}837{,}715$ & 512 & 128 & 4 & 3 & 2 & 0.670 & 160 & 3 \\
   5 & mixed & 180 & 192, 384, 768  & $59{,}006{,}163$ & 512 & 256 & 4 & 4 & 2 & 0.348 & 360 & 3 \\
  \end{tabular}
  \end{ruledtabular}
\end{table*}

\section{Diffusion estimator}\label{app_model}

\subsection{Network and training}\label{app_training}

The denoiser is a U-Net~\cite{ronneberger2015} that predicts $x_0$ directly. The measurement encoder embeds each observable-value pair as a feature vector, or token, retaining the local Pauli symbols and qubit positions. Learned latent vectors collect information from these tokens through attention and are then refined by Transformer layers~\cite{jaegle2021,vaswani2017}, without positional encoding over the observable list. Their average gives the pooled measurement features, which, together with the noise level and the shot count, control channelwise scales and offsets of the U-Net through feature-wise linear modulation~\cite{perez2018}. Bottleneck cross-attention lets the U-Net features at the coarsest resolution draw on the individual latent representations. Table~\ref{tab:arch} specifies the dimensions and training settings.

The observable sampling ratio used in training is bounded below, and a curriculum lowers this bound from $100\%$ to $20\%$ over the initial epochs listed in Table~\ref{tab:arch}. The shot count is drawn uniformly from $\{10^2,10^3,10^4,\infty\}$, where $\infty$ denotes the infinite-shot limit $y_j=\operatorname{Tr}(\rho_0\tilde P_j)$, which retains readout error but has no shot noise. All reported tests use finite shot counts and ratios within the training range.

The PyTorch~\cite{paszke2019} implementation uses AdamW~\cite{loshchilov2019} with minibatches of $512$; $\mu$ and the other network-specific settings in Table~\ref{tab:arch} are chosen on validation data. The networks are trained from random initialization for at most $3500$ epochs, with validation-based learning-rate reduction and early stopping. The reported weights $\theta$ are those with the highest mean fidelity on $512$ fixed validation states over the $15$ finite-shot conditions. No structured-state examples or test data enter training or selection; the weights and $K^*$ are fixed during testing.

\subsection{Number of denoising steps}\label{app_kstar}

Reconstruction uses $K$ evenly spaced indices over the $T$ diffusion steps and the strided posterior update of Appendix~\ref{app_sampler}~\cite{ho2020,nichol2021}. The selected $K^*$ has the highest mean validation fidelity over $240$ states per measurement condition, under two conditions. First, the candidates are $K=2,\ldots,8$, so that every reconstruction includes at least one reverse-process update (Sec.~\ref{sec_method_model}); $K=1$ is evaluated as a separate control (Appendix~\ref{app_dr_steps}). Second, $K^*$ is constrained not to decrease with $n$ within each input-state ensemble. This constraint is a design assumption: the denoised array, with $2d^2$ real entries, grows with $n$, and a larger system is not assigned fewer denoising steps than a smaller one.

\begin{table*}[t]
  \caption{Exploratory reverse-sampler comparison at the validation-selected number of denoising steps $K^*$. Entries are condition-averaged mean fidelities over the $12$ finite-shot conditions with incomplete Pauli sets, averaged over three state-specific initial Gaussian samples $z_T$. Ancestral sampling corresponds to $\kappa=1$ in Eq.~(\ref{eq:transition_noise_scale}), and $\kappa=0$ is the exact strided posterior update with only its injected noise removed. Best $\kappa$ gives the highest mean over the six tested scales, with that scale in parentheses; it is selected on the test data. The last column is ancestral minus DDIM with $\eta_{\mathrm{DDIM}}=0$, computed from unrounded means, with its $95\%$ CI. Because fidelities are averaged over $z_T$, the ancestral column differs slightly from the single-sample \mbox{Diff-QST} means in Table~\ref{tab:dr_macro}.}
  \label{tab:sampler_ablation}
  \begin{ruledtabular}
  \begin{tabular}{cccrrlrl}
  $n$ & Input state & $K^*$ & Ancestral & $\kappa=0$ & Best $\kappa$ & DDIM & Ancestral $-$ DDIM [95\% CI] \\
  \colrule
  2 & pure & 2 & 0.8609 & 0.8647 & 0.8647 ($0.1$) & 0.8602 & $+0.0008\,[-0.0012,+0.0027]$ \\
  3 & pure & 3 & 0.8915 & 0.8960 & 0.8960 ($0$) & 0.8889 & $+0.0026\,[+0.0011,+0.0042]$ \\
  4 & pure & 3 & 0.9302 & 0.9322 & 0.9336 ($0.1$) & 0.9293 & $+0.0010\,[+0.0006,+0.0013]$ \\
  5 & pure & 3 & 0.9075 & 0.9077 & 0.9114 ($0.5$) & 0.9068 & $+0.0007\,[+0.0004,+0.0011]$ \\
  2 & mixed & 2 & 0.8893 & 0.8920 & 0.8920 ($0$) & 0.8893 & $+0.0000\,[-0.0013,+0.0013]$ \\
  3 & mixed & 3 & 0.9101 & 0.9153 & 0.9153 ($0$) & 0.9086 & $+0.0015\,[+0.0002,+0.0027]$ \\
  4 & mixed & 3 & 0.9326 & 0.9291 & 0.9356 ($0.25$) & 0.9313 & $+0.0013\,[+0.0010,+0.0017]$ \\
  5 & mixed & 3 & 0.9185 & 0.8670 & 0.9211 ($0.75$) & 0.9178 & $+0.0006\,[+0.0003,+0.0010]$ \\
  \end{tabular}
  \end{ruledtabular}
\end{table*}

Without the constraint, validation would select the same $K^*$ for seven of the eight networks; for the four-qubit network for mixed input states, the unconstrained maximum occurs at $K=2$, whose fidelity exceeds that at $K=3$ by $1.0\times10^{-4}$. Adding $K=1$ to the candidates leaves all eight values unchanged under the constraint, because $K^*=3$ is still selected at $n=3$. Without the constraint, however, the four-qubit network for mixed input states and the five-qubit networks for pure and mixed input states would select $K=1$, whose validation fidelities exceed those at $K=3$ by $1.9\times10^{-3}$, $5.2\times10^{-3}$, and $2.4\times10^{-3}$, respectively. The constraint therefore does not establish that the unconstrained optimum grows with system size or that the selected value is optimal for each measurement condition.

\subsection{Reverse sampler and initialization}\label{app_sampler}\label{app_initial_noise}

For successive scheduled indices $t'<t$, let $\hat x_0^{(t)}$ be the real representation of the state estimate obtained from $\tilde x_t$. The strided posterior mean and variance are
\begin{align}
 \bar x_{t\to t'}
 &=\frac{\sqrt{\bar\alpha_{t'}}(1-\bar\alpha_t/\bar\alpha_{t'})}{1-\bar\alpha_t}\hat x_0^{(t)}
 \notag\\*
 &\quad+\frac{\sqrt{\bar\alpha_t/\bar\alpha_{t'}}(1-\bar\alpha_{t'})}{1-\bar\alpha_t}\tilde x_t,\\
 \sigma_{t\to t'}^{2}
 &=\frac{(1-\bar\alpha_{t'})(1-\bar\alpha_t/\bar\alpha_{t'})}{1-\bar\alpha_t},
\end{align}
where dependence on $\mathcal M$ and $N_\mathrm{shots}$ is omitted from the notation. Ancestral sampling adds Gaussian noise with this variance to the mean; we refer to it as the injected noise. Updates occur only between scheduled indices, and the final denoiser output is the reported state estimate.

The initial Gaussian sample $z_T$ and the injected noise are separate sources of randomness. We vary each while keeping the networks, $K^*$, test states, measurement outcomes, and observable subsets of the main comparison; the sampler of the main comparison was fixed beforehand and is not changed by these controls. The injected noise is scaled by $\kappa$,
\begin{equation}
 \begin{gathered}
 \tilde x_{t'}=\bar x_{t\to t'}+\kappa\,\sigma_{t\to t'}\,\epsilon,
 \qquad \epsilon\sim\mathcal N(0,I),\\
 \kappa\in\{0,0.1,0.25,0.5,0.75,1\},
 \end{gathered}
 \label{eq:transition_noise_scale}
\end{equation}
so that $\kappa=1$ recovers ancestral sampling and $\kappa=0$ removes the injected noise. The $\kappa=0$ sampler does not yield the Bayesian posterior mean of the density operator and differs from the deterministic sampler of the denoising diffusion implicit model (DDIM), which uses~\cite{song2020ddim}
\begin{equation}
 \begin{aligned}
 \tilde x_{t'}&=\sqrt{\bar\alpha_{t'}}\,\hat x_0^{(t)}+\sqrt{1-\bar\alpha_{t'}}\,\hat\epsilon_t,\\
 \hat\epsilon_t&=\frac{\tilde x_t-\sqrt{\bar\alpha_t}\,\hat x_0^{(t)}}{\sqrt{1-\bar\alpha_t}},
 \end{aligned}
 \label{eq:ddim_initial_control}
\end{equation}
where $\hat\epsilon_t$ is computed from the direct prediction of $x_0$ rather than predicted by a separately trained noise-prediction network. Its stochasticity parameter is $\eta_{\mathrm{DDIM}}=0$, and ancestral sampling is algebraically equivalent to DDIM with $\eta_{\mathrm{DDIM}}=1$ on this schedule~\cite{song2020ddim}. Fidelities are averaged over three replicates of the Gaussian samples and over the $12$ finite-shot conditions with incomplete Pauli sets.

Removing the injected noise ($\kappa=0$) raises the condition-averaged fidelity of six networks, with CIs excluding zero for five of them, but lowers it by $0.0035$ and $0.0515$ for the four- and five-qubit networks for mixed input states (Table~\ref{tab:sampler_ablation}). For the two- and three-qubit networks, no intermediate scale exceeds $\kappa=0$ by more than $2\times10^{-6}$, and the fidelity decreases toward $\kappa=1$. Each four- and five-qubit network instead has an intermediate scale, between $\kappa=0.1$ and $0.75$, whose fidelity exceeds that of the better endpoint by $0.0014$--$0.0037$, with CIs above zero. The best tested scale exceeds ancestral sampling by $0.0026$--$0.0053$; because this scale is selected on the test data, the gain is an optimistic estimate of what tuning $\kappa$ would achieve. For every network, the fidelity range over the tested scales is largest at $r=20\%$, where it is $0.006$--$0.022$, or $0.157$ for five-qubit mixed input states, whose fidelity decreases steadily from $\kappa=0.75$ to $\kappa=0$; the range narrows as $r$ increases, to at most $0.0012$ at $r=100\%$. Ancestral sampling has higher fidelity than deterministic DDIM for the six networks with $n\ge3$, by $0.0006$--$0.0026$ with CIs above zero; the two-qubit CIs include zero. Because DDIM also changes the deterministic update coefficients, this difference does not isolate the effect of the injected noise.

In the main comparison, each state starts from its own Gaussian sample $z_T$. Sharing one standard Gaussian array across all states and conditions changes the condition-averaged fidelity of each network by at most $2.1\times10^{-5}$ at $\kappa=1$ and $\kappa=0$ and by at most $3.5\times10^{-3}$ with deterministic DDIM. The zero array, the mean of $z_T\sim\mathcal N(0,I)$, keeps the starting noise level $\bar\alpha_T$, so it serves as the noisy-sample input rather than as an initial estimate of the density operator. At $\kappa=1$, starting from it changes the fidelity by at most $2.1\times10^{-5}$, except for a decrease of $0.031$ for the two-qubit network for pure input states. At $\kappa=0$, it lowers the fidelity by $0.033$ and $0.048$ for the two- and five-qubit networks for pure input states and changes it by at most $0.002$ for the others; with deterministic DDIM, it raises the fidelity of four networks and lowers that of the other four. Starting from the mean of the initial distribution therefore does not, in general, give the fidelity averaged over Gaussian initial samples. These controls give no reason to replace ancestral sampling from state-specific Gaussian samples in the main comparison: no tested alternative has higher fidelity in all $120$ combinations of network and condition, the best alternative differs between networks, and the largest gain in any incomplete-Pauli condition, $0.017$, is below the smallest incomplete-Pauli fidelity difference from MLE and SDP, $0.021$ (Sec.~\ref{sec_result_accuracy}). For the comparison with the regression baseline, however, the condition-averaged gain of the best tested alternative for each two- and three-qubit network, $0.0027$--$0.0053$ when selected on the test data, is comparable with the magnitudes of their fidelity differences from the regression baseline, $0.0054$--$0.0081$ (Table~\ref{tab:dr_macro}). These comparisons show how sensitive the fidelity is to the sampler settings; they do not test whether the sampled estimates represent calibrated posterior uncertainty.

\section{Statistical and numerical protocol}\label{app_protocol}

\subsection{Paired comparisons}\label{app_stats}

In the main comparison, all estimators receive the same states, observable subsets, and measurement outcomes in each condition. For test state $\nu$, estimator $A$, and subset $\ell$, define $F_{\nu\ell}^{(A)}=F(\rho_{0,\nu},\hat\rho_{0,\nu\ell}^{(A)})$. We first average over subsets and then over states:
\begin{equation}
  \begin{aligned}
  \bar{F}^{(A)}&=\frac{1}{N_\mathrm{test}}\sum_{\nu=1}^{N_\mathrm{test}}
  \bar{F}_{\nu}^{(A)},\\
  \bar{F}_{\nu}^{(A)}&=\frac{1}{L_r}\sum_{\ell=1}^{L_r}F_{\nu\ell}^{(A)}.
  \end{aligned}
  \label{eq:condition_mean_fidelity}
\end{equation}
Here $N_\mathrm{test}=300$ is the number of test states, and $L_r=3$ for $r<100\%$ and $L_r=1$ for the complete Pauli set. For each reference $B\in\{\mathrm{MLE},\mathrm{SDP}\}$, the paired differences are $\bar F_\nu^{(\text{Diff-QST})}-\bar F_\nu^{(B)}$, and their mean is $\bar F^{(\text{Diff-QST})}-\bar F^{(B)}$. Its $95\%$ confidence interval (CI) uses $10{,}000$ percentile bootstrap resamples of the shared state indices, so that all estimators are resampled together~\cite{efron1979,efron1993}; CIs for the mean of a single estimator use the same resampling. These CIs condition on the selected subsets, trained network, and $K^*$. They are pointwise, not simultaneous confidence intervals across the grid. The multiplicity correction in Sec.~\ref{sec_result_accuracy} applies Holm's step-down procedure to two-sided bootstrap $p$-values from $10^6$ resamples per comparison. Other appendices state only departures from this procedure.

Subset dependence is largest relative to the fidelity difference at $n=2$, $r=20\%$, where $m=3$. Averaging over all $455$ three-observable subsets for the same test states and measurement outcomes, \mbox{Diff-QST} retains higher mean fidelity than both references in all six conditions. The differences are $0.004$--$0.010$ for pure and $0.012$--$0.014$ for mixed input states, smaller than for the three main-comparison subsets.
The reversal for the independent Haar-random ensemble at $n=2$, $r=20\%$ (Sec.~\ref{sec_result_shift}) reflects subset choice: applied to the main test states, the three subsets used for that ensemble also give negative differences, and per-subset differences across the two state sets correlate at $0.92$--$0.93$. Averaged over all $455$ subsets, the differences for that ensemble are $0.013$, $0.009$, and $0.006$ at $10^2$, $10^3$, and $10^4$ shots, with both CIs above zero.

\subsection{Reference estimators}\label{app_compute}

MLE starts from the positive-semidefinite projection of the minimum-norm linear least-squares estimate, mixed with $I/d$ at weight $10^{-6}$. With $\delta_\rho$ the Frobenius-norm change and $\delta_{\mathrm{LL}}$ the absolute log-likelihood change per iteration, it stops when $\delta_\rho<10^{-6}$ and either $\delta_{\mathrm{LL}}<10^{-8}$ or $\delta_{\mathrm{LL}}/(mN_\mathrm{shots})<10^{-9}$, or when $\delta_\rho<10^{-7}$, within $2.1\times10^5$ iterations. SDP is solved by SCS~\cite{odonoghue2016,odonoghue2021} through CVXPY~\cite{diamond2016} at tolerance $10^{-8}$. The SDP weights in Eq.~(\ref{eq:sdp}) are
\begin{equation}
 \begin{aligned}
 \hat p_j&=\frac{n_j^++1/2}{N_\mathrm{shots}+1},\\
 v_j&=\frac{4\hat p_j(1-\hat p_j)}{N_\mathrm{shots}},\\
 w_j&=\frac{v_j^{-1}}{m^{-1}\sum_{k=1}^{m}v_k^{-1}},
 \end{aligned}
 \label{eq:sdp_weights}
\end{equation}
where $n_j^+$ is the number of $+1$ outcomes in Eq.~(\ref{eq:measurement_counts}) and $\hat p_j$ is the posterior mean of the outcome probability under the Jeffreys prior; because $0<\hat p_j<1$, each $v_j$ is positive. Outside the early-stopping test of Appendix~\ref{app_timing}, every MLE reconstruction satisfies the stopping criteria except one in the readout test of Appendix~\ref{app_ood_summary}, which reached the iteration limit; for some structured states, the limit is raised. Outside the early-stopping test, every SDP solution is optimal, with one structured-state problem solved by Clarabel~\cite{goulart2026clarabel} instead.

In many incomplete-Pauli conditions, SDP fits the data almost exactly: the optimized measurement-residual term of Eq.~(\ref{eq:sdp}), averaged over states and subsets, is below $10^{-6}$ in $30$ of the $96$ incomplete-Pauli conditions, including every system size and input-state ensemble at $r=20\%$, $10^4$ shots. In such conditions, the regularizer mainly selects among data-consistent states, although it need not select the lowest-purity member of the unregularized solution set.

\subsection{Reconstruction time and early stopping}\label{app_timing}

For timing, MLE and SDP use the same objectives and stopping criteria as in the accuracy comparison, and reference times are mean serial reconstruction times per state. Timings exclude training, network-parameter loading, measurement acquisition, data loading, construction of the effective observables, and fidelity evaluation; GPU times include synchronization and transfer of reconstructed states to the CPU.
For \mbox{Diff-QST} and the regression baseline, time per state is the elapsed time divided by the number of reconstructed states, on the same GPU and software settings. For all eight networks across the five ratios at $10^3$ shots, the regression baseline takes $1.5$--$2.4$~ms per state at batch size $1$, and \mbox{Diff-QST} takes $1.7$--$2.4$ times as long; at batch size $256$, the factor is $1.1$--$1.8$. For four and five qubits over incomplete-Pauli ratios at $10^3$ shots, \mbox{Diff-QST} takes $4.70$--$5.08$~ms per state versus $2.05$--$2.31$~ms for the regression baseline.

Reference times are measured in every nominal shot-ratio condition, whereas each \mbox{Diff-QST} and regression network is timed at every ratio at $10^3$ shots, and that time is used for all shot counts because only input values change. Figure~\ref{fig:time} and Table~\ref{tab:timing_summary} summarize the condition-level ratio $R^{(B)}$ of mean reference time to \mbox{Diff-QST}'s mean time per state, with each of the $30$ conditions at a given system size weighted equally. Averaging the condition-level mean times over the $30$ conditions instead gives ratios of averaged reference time to averaged GPU \mbox{Diff-QST} time of $3.85$, $20.7$, $111$, and $495$ for MLE and $3.79$, $5.53$, $21.7$, and $127$ for SDP at $n=2,3,4,5$, respectively. At batch size $256$, \mbox{Diff-QST}'s GPU time per state is shorter than the mean MLE and SDP times in every nominal condition.

\begin{table*}[t]
\caption{Median reference/\mbox{Diff-QST} reconstruction-time ratios at batch size $1$, with minima in parentheses, across $30$ equally weighted conditions per system size. MLE and SDP use the $32$-core CPU configuration; \mbox{Diff-QST} runs on the GPU or uses $32$ CPU cores as indicated. A minimum above one means shorter \mbox{Diff-QST} mean times throughout the evaluated conditions.}
\label{tab:timing_summary}
\begin{ruledtabular}
\begin{tabular}{ccccc}
 & \multicolumn{2}{c}{\mbox{Diff-QST} on GPU} & \multicolumn{2}{c}{\mbox{Diff-QST} on CPU} \\
\cmidrule(r{2pt}){2-3}\cmidrule(l{2pt}){4-5}
$n$ & MLE/\mbox{Diff-QST} & SDP/\mbox{Diff-QST} & MLE/\mbox{Diff-QST} & SDP/\mbox{Diff-QST} \\
\colrule
$2$ & $1.55\;(0.107)$ & $3.67\;(3.07)$ & $0.977\;(0.0584)$ & $2.63\;(1.68)$ \\
$3$ & $9.76\;(0.449)$ & $5.53\;(2.80)$ & $1.74\;(0.0829)$ & $0.963\;(0.518)$ \\
$4$ & $70.8\;(14.9)$ & $21.1\;(13.6)$ & $10.2\;(2.25)$ & $2.98\;(2.04)$ \\
$5$ & $482\;(135)$ & $125\;(71.9)$ & $31.4\;(9.22)$ & $8.25\;(4.83)$ \\
\end{tabular}
\end{ruledtabular}
\end{table*}

On the $32$-core CPU configuration at batch size $1$, \mbox{Diff-QST} takes $31.04$--$35.51$~ms per state at four qubits and $74.57$--$78.23$~ms at five. It has lower condition-averaged time than MLE at all four sizes and than SDP at two, four, and five qubits; at three qubits, SDP's condition-averaged time is $0.971$ times that of \mbox{Diff-QST}. These results describe the tested implementations and hardware, not hardware-independent speedups.

MLE and SDP times vary across test states, as in the five-qubit example of Sec.~\ref{sec_result_cost}. Across the five-qubit grid, MLE mean times and their spread across states are largest at $r=80\%$; for pure input states at $10^3$ shots and this ratio, one reconstruction takes $16.8$~s. SDP varies less but has occasional long reconstructions: for pure input states at $10^4$ shots and $r=40\%$, its iteration counts range from $250$ to $5575$, and one reconstruction takes $6.06$~s. The per-state reference times also include fluctuations from a shared server, so their spread does not isolate the contribution of solver iterations.

\begin{figure*}[t]
  \centering
  \includegraphics[width=\textwidth]{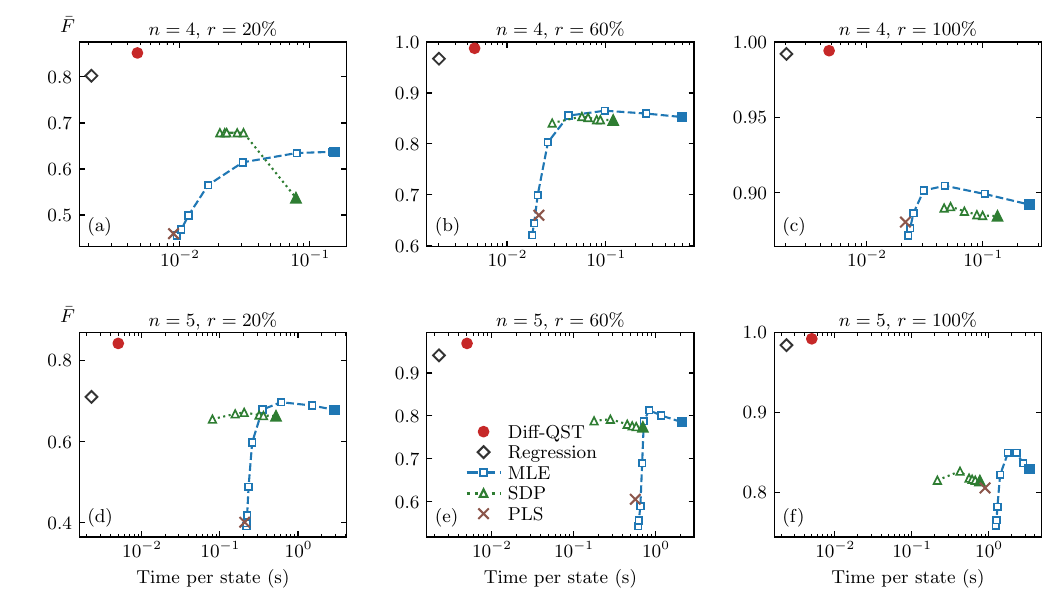}
  \caption{Mean fidelity versus reconstruction time per state for pure input states at $N_\mathrm{shots}=10^3$, (a)--(c) $n=4$ and (d)--(f) $n=5$ at $r=20\%$, $60\%$, and $100\%$. Filled circles show \mbox{Diff-QST} and open diamonds the regression baseline, both at their GPU times per state at batch size $1$; crosses show projected least squares (PLS). Open squares show MLE stopped after $0$ to $3\times10^4$ iterations and filled squares the converged MLE; open triangles show SDP at SCS tolerances from $10^{-2}$ to $10^{-6}$ and filled triangles at $10^{-8}$. All times are measured on the $15$ states and observable subsets of the reconstruction-time comparison. Vertical scales differ between panels. Mixed input states, which complete the $12$ conditions of the early-stopping comparison, are not shown.}
  \label{fig:accuracy_time}
\end{figure*}

For both input-state ensembles at $n=4$ and $5$ and at $N_\mathrm{shots}=10^3$ and $r=20\%$, $60\%$, and $100\%$, we also stop MLE after $0$, $10$, $30$, $100$, $300$, $10^3$, $3\times10^3$, $10^4$, or $3\times10^4$ iterations and solve SDP at SCS tolerances from $10^{-2}$ to $10^{-8}$; iteration $0$ is the MLE starting estimate. MLE times are the elapsed times at each iteration limit within one run per state, and each SDP tolerance is a separate solve; all are measured on the $15$ states and observable subsets of the CPU reference timing (Fig.~\ref{fig:accuracy_time}). Projected least squares (PLS), the density operator closest in Frobenius norm to the least-squares estimate~\cite{smolin2012}, is included as a noniterative estimator.

Early stopping can raise the mean fidelity above the converged value, by up to $0.045$ for MLE and $0.142$ for SDP, but in no condition to that of \mbox{Diff-QST} (Sec.~\ref{sec_result_cost}). Stopping each estimator when its mean fidelity first comes within $0.001$ of its converged value leaves MLE and SDP $15$--$676$ and $13$--$145$ times slower than GPU \mbox{Diff-QST}. No tested stopping point of MLE or SDP, including iteration $0$ and SCS tolerance $10^{-2}$, is faster than GPU \mbox{Diff-QST} in any of the $12$ conditions. PLS has mean fidelity $0.40$--$0.88$, and its time depends on the least-squares implementation. Faster MLE and SDP algorithms (Sec.~\ref{sec_result_cost}) would reach the same converged estimates wherever these are unique: for SDP, whose objective is strictly convex, and for MLE with the complete Pauli set.

\section{Further accuracy comparisons}\label{app_paired}

\begin{table*}[t]
  \caption{Sensitivity of the comparison to the reference estimator and the accuracy measure. The HMLE columns cover the $12$ finite-shot conditions with incomplete Pauli sets: the condition-averaged mean fidelity of HMLE, the difference $\bar F^{(\text{Diff-QST})}-\bar F^{(\mathrm{HMLE})}$ computed from unrounded means with its $95\%$ CI, and the number of conditions with a positive condition-level difference. The last column gives the median and, in brackets, the maximum over the $15$ shot-ratio conditions of $\Delta\bar D_{\rm tr}$, the trace-distance difference of \mbox{Diff-QST} from the closer of MLE and SDP; negative values favor \mbox{Diff-QST}. The \mbox{Diff-QST} mean fidelities are those in the $K^*$ column of Table~\ref{tab:dr_macro}.}
  \label{tab:reference_metric}
  \begin{ruledtabular}
  \begin{tabular}{cclcc}
   & \multicolumn{3}{c}{Hedged MLE} & Trace distance \\
  \cmidrule(r{2pt}){2-4}\cmidrule(l{2pt}){5-5}
  $n$ & $\bar F^{(\mathrm{HMLE})}$ & \mbox{Diff-QST} $-$ HMLE [95\% CI] & Positive & $\Delta\bar D_{\rm tr}$ median [max] \\
  \colrule
  \multicolumn{5}{l}{Pure input states} \\
  2 & 0.8440 & $+0.0173\,[+0.0138,+0.0207]$ & $9/12$ & $-0.0398\;[+0.0066]$ \\
  3 & 0.8504 & $+0.0421\,[+0.0398,+0.0444]$ & $12/12$ & $-0.0681\;[+0.0057]$ \\
  4 & 0.8514 & $+0.0789\,[+0.0780,+0.0798]$ & $12/12$ & $-0.0879\;[-0.0184]$ \\
  5 & 0.8021 & $+0.1058\,[+0.1051,+0.1066]$ & $12/12$ & $-0.0817\;[+0.0184]$ \\
  \colrule
  \multicolumn{5}{l}{Mixed input states} \\
  2 & 0.8751 & $+0.0140\,[+0.0120,+0.0160]$ & $10/12$ & $-0.0497\;[+0.0032]$ \\
  3 & 0.8727 & $+0.0391\,[+0.0374,+0.0408]$ & $12/12$ & $-0.1054\;[-0.0038]$ \\
  4 & 0.8539 & $+0.0787\,[+0.0777,+0.0797]$ & $12/12$ & $-0.1455\;[-0.0385]$ \\
  5 & 0.7958 & $+0.1221\,[+0.1205,+0.1236]$ & $12/12$ & $-0.1540\;[-0.0756]$ \\
  \end{tabular}
  \end{ruledtabular}
\end{table*}

\subsection{Observables per state parameter and ratios below the training range}\label{app_coverage}

\begin{figure*}[t]
  \centering
  \includegraphics[width=\textwidth]{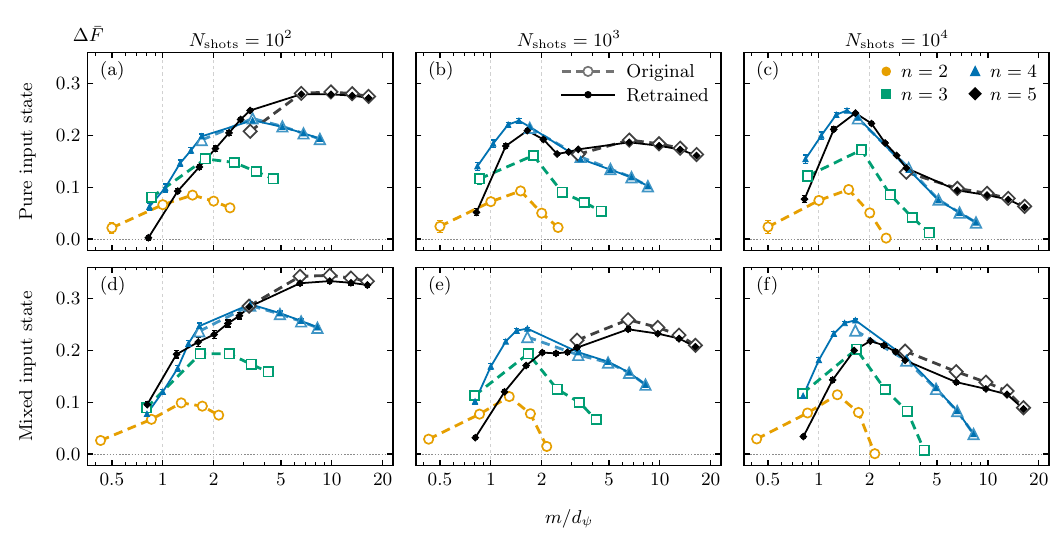}
  \caption{Fidelity difference $\Delta\bar F=\bar F^{(\text{Diff-QST})}-\max\{\bar F^{(\mathrm{MLE})},\bar F^{(\mathrm{SDP})}\}$ versus $m/d_\psi$, the number of measured observables per real parameter of the input-state family, for (a)--(c) pure and (d)--(f) mixed input states at $N_{\mathrm{shots}}=10^2,10^3,10^4$; along each curve, $n$ is fixed and $r$ varies. Original networks (open markers) were trained with observable sampling ratios down to $20\%$; retrained networks (filled markers), down to $10\%$ for $n=4$ and $5\%$ for $n=5$. Most error bars, $95\%$ CIs relative to the better of MLE and SDP, are smaller than the markers. Vertical lines mark $m=d_\psi$, the dimensional lower bound on the number of measured quantities for identifying every member of the input-state family, and $m=2d_\psi$, about the number that topological obstructions can require~\cite{heinosaari2013}; the horizontal line marks $\Delta\bar F=0$.}
  \label{fig:margin}
\end{figure*}

Let $d_\psi$ denote the number of real parameters of the input-state family, $d_\psi=2^{n+1}-2$ for pure and $2^{n+1}-1$ for mixed input states. Expressing the measurement count as $m/d_\psi$ distinguishes the fraction of the complete Pauli set from the number of observables available per state parameter (Fig.~\ref{fig:margin}). The fixed nominal preparation channel acts as an invertible linear map on density operators, so each target family has the same dimension $d_\psi$ as its input-state family.

To determine whether the location of the largest fidelity difference scales with $d_\psi$ below the $20\%$ lower bound of the training ratio, we retrained the four- and five-qubit networks with this bound reduced to $10\%$ and $5\%$, respectively. The retrained networks retain the architecture and training settings of the corresponding networks in Table~\ref{tab:arch} and use $K^*=3$; at $n=4$, the validation maxima at $K=2$, within $1.1\times10^{-3}$ of $K=3$, are excluded by the nondecreasing constraint of Appendix~\ref{app_kstar}. They are evaluated on the $300$ test states of the main comparison at the nominal ratios and at $r=10\%$ to $17.5\%$ for $n=4$ and $r=5\%$ to $17.5\%$ for $n=5$, in steps of $2.5\%$, with MLE and SDP computed from the same observable subsets and measurement outcomes. These networks are used only for this test.

For the retrained networks, both CIs lie above zero in $119$ of their $120$ conditions; at $n=5$, pure input states, $10^2$ shots, and $r=5\%$, the positive mean difference has a CI relative to MLE that includes zero. Retraining changes fidelity in both directions at $r=20\%$; at $r\ge40\%$, the mean fidelities of the retrained networks lie within $0.006$ of those of the original networks, except for the five-qubit network for mixed input states, whose means are $0.002$--$0.020$ lower.

The exception to the range $m/d_\psi=1.47$--$2.02$ given in Sec.~\ref{sec_result_accuracy} for $N_{\mathrm{shots}}\ge10^3$ is five-qubit mixed input states at $10^3$ shots, for which the largest observed fidelity difference occurs at $m/d_\psi=6.49$.
At $10^2$ shots per observable, all four retrained networks attain their largest observed fidelity differences at higher ratios within the evaluated grid, $m/d_\psi=3.29$--$9.73$. This shift is consistent with larger shot noise weakening the constraints supplied by each observable, so that more measurements are needed before the references narrow the fidelity difference. The location of the largest observed fidelity difference on the $m/d_\psi$ axis therefore depends on $N_\mathrm{shots}$.

From the observed maxima to the complete Pauli set, the reference estimators show larger overall fidelity increases than \mbox{Diff-QST}, narrowing the fidelity difference. This decrease does not imply that \mbox{Diff-QST} has already saturated. For example, for five-qubit pure input states at $10^4$ shots, increasing the number of measured observables from $m=1.65d_\psi$ to the complete Pauli set raises the retrained network's mean fidelity from $0.707$ to $0.993$, while MLE and SDP improve from $0.464$ to $0.932$ and from $0.363$ to $0.923$, respectively. In this example, the smallest evaluated measurement counts attaining $\bar F\ge0.90$ are $m=409$ for \mbox{Diff-QST} and $m=818$ for both references.

\begin{table*}[t]
  \caption{Mean fidelity difference $\bar F^{(\text{Diff-QST})}-\bar F^{(\mathrm{Reg})}$ from the regression baseline by observable sampling ratio. Each entry equally weights the three finite shot counts. Positive values favor \mbox{Diff-QST}. A dagger marks the entries whose $95\%$ CI, obtained by resampling states after averaging over the three shot counts, includes zero; all other CIs exclude zero.}
  \label{tab:dr_ratio}
  \begin{ruledtabular}
  \begin{tabular}{ccrrrrr}
  $n$ & Input state & $20\%$ & $40\%$ & $60\%$ & $80\%$ & $100\%$ \\
  \colrule
  2 & pure & $-0.0254$ & $-0.0052$\rlap{$^{\dagger}$} & $-0.0009$\rlap{$^{\dagger}$} & $+0.0008$ & $+0.0002$ \\
  3 & pure & $-0.0274$ & $-0.0004$\rlap{$^{\dagger}$} & $-0.0023$ & $-0.0024$ & $-0.0022$ \\
  4 & pure & $+0.0424$ & $+0.0369$ & $+0.0182$ & $+0.0073$ & $+0.0008$ \\
  5 & pure & $+0.1343$ & $+0.0723$ & $+0.0392$ & $+0.0241$ & $+0.0141$ \\
  \colrule
  2 & mixed & $-0.0108$ & $-0.0067$ & $-0.0060$ & $-0.0011$ & $-0.0004$ \\
  3 & mixed & $-0.0185$ & $+0.0002$\rlap{$^{\dagger}$} & $-0.0019$ & $-0.0016$ & $-0.0013$ \\
  4 & mixed & $+0.0284$ & $+0.0338$ & $+0.0180$ & $+0.0077$ & $+0.0007$ \\
  5 & mixed & $+0.0380$ & $+0.0221$ & $+0.0125$ & $+0.0085$ & $+0.0051$ \\
  \end{tabular}
  \end{ruledtabular}
\end{table*}

\begin{table*}[t]
  \caption{Additional training runs for the comparison with the regression baseline. Each system size and input-state ensemble has three further runs per estimator, paired by initialization seed: $1001$, $2001$, and $3001$ at $n=2,3,4$ and $51001$, $51002$, and $51003$ at $n=5$. Entries are mean fidelities equally weighting the same $12$ finite-shot conditions with incomplete Pauli sets as Table~\ref{tab:dr_macro}; \mbox{Diff-QST} uses $K^*=2$ at $n=2$ and $K^*=3$ otherwise. $\Delta=\bar F^{(K^*)}-\bar F^{(\mathrm{Reg})}$ is computed from unrounded means, and positive values favor \mbox{Diff-QST}. All $24$ differences have $95\%$ CIs that exclude zero, with half-widths $0.0009$--$0.0039$; the CIs use $10{,}000$ paired bootstrap resamples of the $300$ test states after averaging subsets and conditions, and each is conditional on the corresponding pair of trained networks. The reported networks in Table~\ref{tab:dr_macro} are not repeated here.}
  \label{tab:dr_seeds}
  \begin{ruledtabular}
  \begin{tabular}{crrrrrrrrr}
   & \multicolumn{3}{c}{Run 1} & \multicolumn{3}{c}{Run 2} & \multicolumn{3}{c}{Run 3} \\
  \cmidrule(lr){2-4}\cmidrule(lr){5-7}\cmidrule(lr){8-10}
  $n$ & Regression & \mbox{Diff-QST} & $\Delta$ & Regression & \mbox{Diff-QST} & $\Delta$ & Regression & \mbox{Diff-QST} & $\Delta$ \\
  \colrule
  \multicolumn{10}{l}{Pure input states} \\
  2 & $0.8697$ & $0.8579$ & $-0.0118$ & $0.8690$ & $0.8562$ & $-0.0128$ & $0.8700$ & $0.8623$ & $-0.0077$ \\
  3 & $0.8972$ & $0.8924$ & $-0.0049$ & $0.8992$ & $0.8940$ & $-0.0052$ & $0.8990$ & $0.8926$ & $-0.0064$ \\
  4 & $0.8994$ & $0.9313$ & $+0.0319$ & $0.9083$ & $0.8986$ & $-0.0097$ & $0.9015$ & $0.9410$ & $+0.0396$ \\
  5 & $0.8986$ & $0.9190$ & $+0.0204$ & $0.9084$ & $0.9177$ & $+0.0093$ & $0.8816$ & $0.9189$ & $+0.0373$ \\
  \colrule
  \multicolumn{10}{l}{Mixed input states} \\
  2 & $0.8953$ & $0.8861$ & $-0.0092$ & $0.8946$ & $0.8873$ & $-0.0073$ & $0.8950$ & $0.8882$ & $-0.0068$ \\
  3 & $0.9178$ & $0.9092$ & $-0.0086$ & $0.9166$ & $0.9048$ & $-0.0119$ & $0.9166$ & $0.9094$ & $-0.0072$ \\
  4 & $0.9208$ & $0.9381$ & $+0.0173$ & $0.9098$ & $0.9392$ & $+0.0294$ & $0.9109$ & $0.9032$ & $-0.0077$ \\
  5 & $0.8639$ & $0.9267$ & $+0.0628$ & $0.8971$ & $0.9141$ & $+0.0169$ & $0.9019$ & $0.9237$ & $+0.0217$ \\
  \end{tabular}
  \end{ruledtabular}
\end{table*}

\subsection{Hedged maximum-likelihood reference}\label{app_hmle}

Hedged MLE (HMLE) maximizes $L(\rho)\,(\det\rho)^{1/2}$~\cite{blumekohout2010hedged}, where $L(\rho)$ is the binary likelihood used by MLE. The determinant term enforces full rank, and the exponent $1/2$ gives the add-$1/2$ (Jeffreys) frequency rule for an ideal projective measurement in one orthonormal basis. HMLE uses the same starting state and stopping criteria as MLE, and all its reconstructions satisfy them. Table~\ref{tab:reference_metric} summarizes the incomplete-Pauli comparison; $\Delta\bar F$ in the main text remains relative to the better of MLE and SDP.

Relative to HMLE, \mbox{Diff-QST} has higher mean fidelity in $91$ of the $96$ incomplete-Pauli conditions and all $24$ complete-Pauli conditions; the CIs lie above zero in $89$ and $24$, respectively, and none lies entirely below zero. The five reversals occur at $n=2$, $r=20\%$, for pure input states at all shot counts and mixed input states at $10^2$ and $10^3$ shots, with differences from $-0.0076$ to $-0.0013$.

\subsection{Trace distance}\label{app_distance}

The trace distance $D_{\rm tr}$ of Sec.~\ref{sec_result_accuracy} is evaluated on the reconstructions of the main comparison and enters neither training nor selection. Mean distances to $\rho_0$ are formed as in Eq.~(\ref{eq:condition_mean_fidelity}); negative differences favor \mbox{Diff-QST}. The last column of Table~\ref{tab:reference_metric} summarizes $\Delta\bar D_{\rm tr}=\bar D_{\rm tr}^{(\text{Diff-QST})}-\min\{\bar D_{\rm tr}^{(\mathrm{MLE})},\bar D_{\rm tr}^{(\mathrm{SDP})}\}$, the difference from the reference with the smaller mean distance.

The $112$ conditions in which \mbox{Diff-QST} is closer to the target (Sec.~\ref{sec_result_accuracy}) include $91$ of the $96$ incomplete-Pauli conditions. Three of the four in which the references are closer use the complete Pauli set at $10^4$ shots, for two- and three-qubit pure and two-qubit mixed input states, where the fidelity difference is also smallest (Sec.~\ref{sec_result_accuracy}). The fourth is five-qubit pure input states at $10^4$ shots and $r=40\%$. There, the target states are full rank, whereas the MLE and SDP estimates are often rank deficient~\cite{blumekohout2010hedged}: on average, $13.3$ and $19.2$ of their $32$ eigenvalues lie below $10^{-4}$, compared with $3.2$ for \mbox{Diff-QST}. Because fidelity depends on the square roots of the eigenvalues, these near-zero eigenvalues affect fidelity more strongly than trace distance. The dominant eigenvectors of the MLE and SDP estimates also overlap more with those of the targets, with mean squared overlap $0.998$ for both, compared with $0.949$ for \mbox{Diff-QST}. In the remaining four conditions, all at $n=2$ and $r=20\%$, both CIs include zero.

\begin{table*}[t]
\caption{Smallest evaluated total shot budget attaining the specified mean fidelity at five qubits, over the expanded $21$-allocation grid per input-state ensemble. $m$ is the number of Pauli observables and $N_{\rm shots}$ the shots per observable. The attained $\bar F$ and its $95\%$ CI over test states describe the selected condition, not uncertainty in the minimum budget or its ratio between estimators.}
\label{tab:dr_budget}
\begin{ruledtabular}
\begin{tabular}{cllrrrrl}
Target & Input state & Estimator & $m$ & $N_{\rm shots}$ & $N_{\rm tot}$ & $\bar F$ & $95\%$ CI \\
\colrule
0.90 & pure & \mbox{Diff-QST} & $613$ & $100$ & $61{,}300$ & $0.9145$ & $[0.9134,0.9157]$ \\
0.90 & pure & Regression & $1023$ & $100$ & $102{,}300$ & $0.9292$ & $[0.9272,0.9312]$ \\
0.90 & mixed & \mbox{Diff-QST} & $613$ & $100$ & $61{,}300$ & $0.9090$ & $[0.9076,0.9104]$ \\
0.90 & mixed & Regression & $818$ & $100$ & $81{,}800$ & $0.9287$ & $[0.9272,0.9301]$ \\
0.95 & pure & \mbox{Diff-QST} & $1023$ & $100$ & $102{,}300$ & $0.9583$ & $[0.9575,0.9590]$ \\
0.95 & pure & Regression & $1023$ & $200$ & $204{,}600$ & $0.9581$ & $[0.9570,0.9592]$ \\
0.95 & mixed & \mbox{Diff-QST} & $1023$ & $100$ & $102{,}300$ & $0.9525$ & $[0.9516,0.9534]$ \\
0.95 & mixed & Regression & $613$ & $500$ & $306{,}500$ & $0.9547$ & $[0.9539,0.9555]$ \\
\end{tabular}
\end{ruledtabular}
\end{table*}

\begin{figure*}[t]
\centering
\includegraphics[width=7.05in]{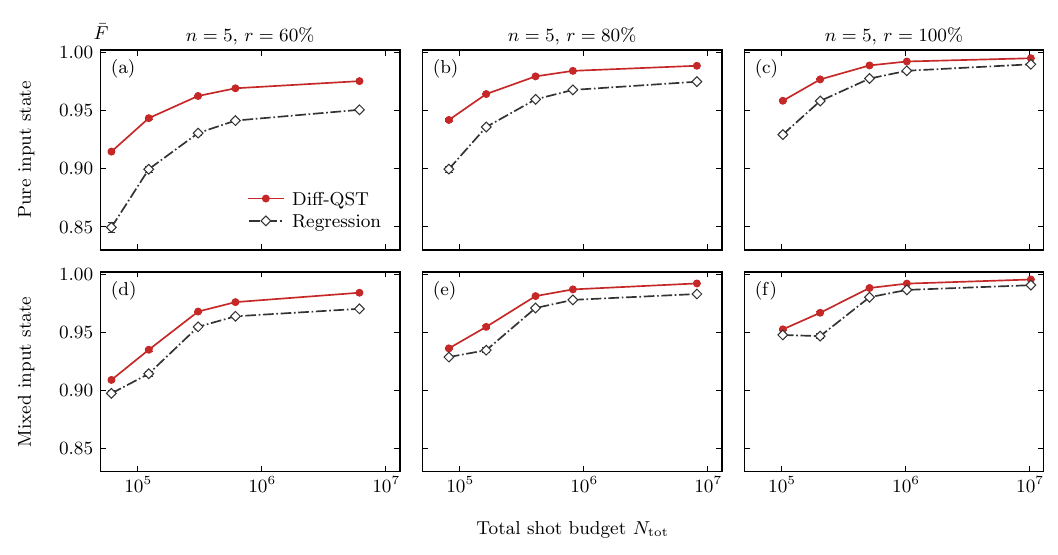}
\caption{Five-qubit condition-level mean fidelity of \mbox{Diff-QST} and the regression baseline versus total shot budget $N_{\rm tot}=mN_{\rm shots}$ for (a)--(c) pure and (d)--(f) mixed input states at $r=60\%,80\%,100\%$ ($m=613,818,1023$). Each curve contains $N_{\rm shots}=100,200,500,1000,10000$. Symbols give means over $300$ test states; bars are $95\%$ bootstrap CIs over test states. Connecting lines guide the eye.}
\label{fig:dr_budget}
\end{figure*}

\section{Further comparisons with the regression baseline}\label{app_dr}

For the condition-averaged comparisons in Table~\ref{tab:dr_macro}, CIs resample state indices after averaging subsets and conditions. The shared learning rate, ratio curriculum, and infidelity weight $\mu$ are listed in Table~\ref{tab:arch}.

At four and five qubits, the fidelity differences are larger at low observable sampling ratios; all $16$ shot-averaged four- and five-qubit differences for incomplete Pauli sets have CIs above zero. At $r=20\%$, all four- and five-qubit networks have higher mean fidelity than the regression baseline at every shot count, with CIs above zero.
Adding the three complete-Pauli conditions preserves all eight directions in Table~\ref{tab:dr_macro}, with CIs still excluding zero. Table~\ref{tab:dr_ratio} averages over shot counts within each ratio; $36$ of its $40$ CIs exclude zero. At $10^2$ shots and $r=100\%$, the regression baseline has slightly higher mean fidelity for both four-qubit networks.

Table~\ref{tab:dr_seeds} reports the three additional training runs per estimator for each system size and input-state ensemble, trained from further random initializations with unchanged settings; Sec.~\ref{sec_result_controls} summarizes their signs. In the two four-qubit pairs that favor the regression baseline, by $0.010$ for pure and $0.008$ for mixed input states, the \mbox{Diff-QST} network has a lower condition-averaged mean fidelity, $0.899$ and $0.903$, than the other three \mbox{Diff-QST} networks, $0.930$--$0.941$.

\begin{table*}[t]
  \caption{Mean fidelity difference $\bar F^{(\text{Diff-QST})}-\bar F^{(\mathrm{Reg})}$ from the regression baseline under the distribution shifts of Sec.~\ref{sec_result_shift}. Each entry averages the condition-level differences over the conditions of its row at one system size, whose number is given by Cond.; parentheses give the numbers of these conditions whose $95\%$ CI lies above and below zero. Positive values favor \mbox{Diff-QST}. Nominal rows use the $15$ finite-shot conditions of the main comparison, and preparation-noise rows use $r=20\%$, $60\%$, and $100\%$. Structured rows pool the canonical representatives or ensembles named in the first column.}
  \label{tab:dr_shift}
  \begin{ruledtabular}
  \begin{tabular}{llrllll}
  Change & Input states & Cond. & \multicolumn{1}{c}{$n=2$} & \multicolumn{1}{c}{$n=3$} & \multicolumn{1}{c}{$n=4$} & \multicolumn{1}{c}{$n=5$} \\
  \colrule
  Nominal & pure & $15$ & $-0.0061$ (4/5) & $-0.0069$ (1/12) & $+0.0211$ (14/1) & $+0.0568$ (15/0) \\
  Nominal & mixed & $15$ & $-0.0050$ (2/11) & $-0.0046$ (4/9) & $+0.0177$ (14/1) & $+0.0172$ (15/0) \\
  \colrule
  Preparation $\times0.25$ & pure & $9$ & $+0.0043$ (5/0) & $+0.0030$ (5/1) & $+0.0277$ (8/1) & $+0.0753$ (9/0) \\
  Preparation $\times0.25$ & mixed & $9$ & $-0.0053$ (2/6) & $-0.0037$ (1/6) & $+0.0181$ (7/1) & $+0.0281$ (8/0) \\
  Preparation $\times0.5$ & pure & $9$ & $-0.0010$ (3/0) & $-0.0028$ (4/3) & $+0.0261$ (8/1) & $+0.0729$ (9/0) \\
  Preparation $\times0.5$ & mixed & $9$ & $-0.0059$ (2/7) & $-0.0050$ (1/7) & $+0.0183$ (8/1) & $+0.0235$ (9/0) \\
  Readout $\times0.5$ & pure & $15$ & $-0.0069$ (4/6) & $-0.0069$ (2/12) & $+0.0207$ (14/1) & $+0.0549$ (15/0) \\
  Readout $\times0.5$ & mixed & $15$ & $-0.0050$ (1/11) & $-0.0054$ (2/12) & $+0.0179$ (13/1) & $+0.0184$ (15/0) \\
  Readout $\times2$ & pure & $15$ & $-0.0046$ (4/3) & $-0.0092$ (0/13) & $+0.0175$ (14/1) & $+0.0617$ (15/0) \\
  Readout $\times2$ & mixed & $15$ & $-0.0053$ (2/11) & $-0.0035$ (6/7) & $+0.0186$ (14/0) & $+0.0221$ (15/0) \\
  Degeneracy lifted & mixed & $15$ & $-0.0062$ (2/12) & $-0.0043$ (4/10) & $+0.0189$ (14/1) & $+0.0166$ (15/0) \\
  Lower purity range & mixed & $15$ & $-0.0033$ (4/9) & $-0.0105$ (0/15) & $+0.0051$ (10/3) & $+0.0101$ (15/0) \\
  \colrule
  Canonical GHZ & pure & $45$ & $-0.0395$ (6/26) & $-0.0148$ (0/19) & $-0.0253$ (0/39) & $-0.0292$ (1/34) \\
  Canonical W, cluster & pure & $30$ & $+0.0050$ (11/3) & $-0.0275$ (0/24) & $+0.0067$ (20/1) & $+0.0289$ (29/0) \\
  Rotated ensembles & pure & $45$ & $-0.0140$ (10/19) & $-0.0110$ (2/34) & $+0.0195$ (41/3) & $+0.0599$ (45/0) \\
  Haar-random ensemble & pure & $15$ & $-0.0068$ (3/4) & $-0.0082$ (2/12) & $+0.0219$ (14/1) & $+0.0563$ (15/0) \\
  \end{tabular}
  \end{ruledtabular}
\end{table*}

\subsection{One- and three-step reconstruction}\label{app_dr_steps}

We evaluate the trained networks at $K=1$ and the selected $K^*=3$ for $n=3,4,5$ in the $15$ finite-shot conditions of the main comparison and, for the five-qubit networks, the six additional conditions of Appendix~\ref{app_dr_shots}. No weights are retrained, and $K^*$ is not reselected. The two-qubit networks, whose selected value is $K^*=2$, are outside this control. For $K=1$, the denoiser is applied once at $\bar\alpha_T$, the noise level at which reconstruction starts, to the initial Gaussian sample $z_T$, and its state estimate is the reconstruction; the $K=3$ reconstruction uses the same initial sample.

In the $K=1$ columns of Table~\ref{tab:dr_macro}, $\bar F^{(1)}-\bar F^{(\mathrm{Reg})}$ compares one network evaluation per estimator, and $\bar F^{(K^*)}-\bar F^{(1)}$ compares the two reconstruction procedures at fixed diffusion-network parameters. The $K=1$ control is not evaluated under the distribution shifts of Appendix~\ref{app_dr_shift}.

\subsection{Total shot budgets at specified fidelity}\label{app_dr_shots}

We extend the five-qubit test to $N_{\rm shots}=200,500$ at $r=60\%,80\%,100\%$, retaining the trained networks, $K^*=3$, and the $300$ targets of the main comparison for each input-state ensemble. This exploratory extension adds six conditions per ensemble to the $15$-condition finite-shot grid of the main comparison, giving $21$ evaluated allocations, that is, combinations of $m$ and $N_{\rm shots}$, per estimator and ensemble.
At each specified fidelity, Table~\ref{tab:dr_budget} selects the smallest evaluated $N_{\rm tot}$ whose condition-level mean fidelity meets the target. Figure~\ref{fig:dr_budget} shows the dependence on $N_{\rm tot}$ for the three ratios evaluated at the additional shot counts.

At target $0.95$, both \mbox{Diff-QST} allocations use the complete Pauli set, so the reductions given in Sec.~\ref{sec_result_controls} concern total shots, not fewer measured observables. At target $0.90$, all four selected allocations use $100$ shots per observable, so the reductions apply to both total shots and measured observables. Using the lower CI endpoints instead of the means selects the same budgets. In all $12$ additional conditions, \mbox{Diff-QST} has higher mean fidelity than the regression baseline, with CIs above zero and differences of $0.0080$--$0.0439$.

The $K=1$ control reaches mean fidelity $0.90$ at $N_{\rm tot}=61{,}300$ and $0.95$ at $N_{\rm tot}=102{,}300$ in both input-state ensembles, the same smallest evaluated budgets as $K^*=3$, so the reductions relative to the regression baseline persist without repeated denoising. The allocation grid is nonuniform: $200$ and $500$ shots are not tested at $r=20\%$ and $40\%$.

\subsection{Distribution shifts}\label{app_dr_shift}

We apply the trained regression networks, without retraining, to the distribution-shift tests of Sec.~\ref{sec_result_shift}. Table~\ref{tab:dr_shift} gives the fidelity differences of \mbox{Diff-QST} from the regression baseline by system size; for the structured states, the CIs resample acquisitions and states as in Appendix~\ref{app_structured}.

At four and five qubits, the CI lies above zero in $235$ of the $238$ spectral, readout, and preparation-noise conditions in which \mbox{Diff-QST} has the higher mean fidelity (Sec.~\ref{sec_result_shift}). For pure input states, the difference grows under weaker preparation noise: at five qubits it is $0.075$ and $0.073$ at quarter and half strength, compared with $0.063$ for the same shot counts and ratios under nominal noise. It shrinks in the lower purity range. At two and three qubits, the regression baseline retains its nominal advantage under the readout and mixed-state distribution changes and under weaker preparation noise for mixed input states; for pure input states, its advantage reverses at quarter preparation strength and shrinks at half strength.

Under weaker preparation noise, the regression baseline also reverses relative to MLE and SDP more often than \mbox{Diff-QST} (Sec.~\ref{sec_result_shift}); at $n=5$, quarter strength, $10^4$ shots, and $r=20\%$, the mean fidelities are $0.648$, $0.798$, and $0.899$ for the regression baseline, \mbox{Diff-QST}, and MLE. Under the readout and mixed-state distribution changes, the numbers of conditions with both CIs above zero differ from those of \mbox{Diff-QST} by at most two. In the $39$ two- and three-qubit structured conditions in which both CIs of \mbox{Diff-QST} lie below zero, the regression baseline has higher mean fidelity than \mbox{Diff-QST} in all $39$, with CIs excluding zero in $34$, and exceeds the mean fidelities of both MLE and SDP in $22$.

\section{Distribution-shift tests}\label{app_shift}

\begin{table*}[t]
  \caption{Canonical-state fidelity differences of \mbox{Diff-QST} from the better of MLE and SDP, summarized by the median over $15$ measurement conditions. Parentheses give the numbers of conditions in which both CIs lie above zero and below zero, respectively. The final column gives the purity of the five-qubit target.}
  \label{tab:structured}
  \begin{ruledtabular}
  \begin{tabular}{lccccc}
  State & $n=2$ & $n=3$ & $n=4$ & $n=5$ & $\operatorname{Tr}\rho_0^{2}$ \\
  \colrule
  GHZ$^{+}$ & $-0.058$ (4, 8) & $+0.046$ (12, 2) & $+0.069$ (13, 0) & $+0.115$ (15, 0) & $0.566$ \\
  GHZ$^{-}$ & $-0.018$ (5, 5) & $+0.041$ (9, 2) & $+0.078$ (12, 0) & $+0.107$ (15, 0) & $0.566$ \\
  GHZ$^{i}$ & $-0.008$ (6, 7) & $+0.060$ (12, 0) & $+0.090$ (13, 0) & $+0.110$ (15, 0) & $0.566$ \\
  W & $+0.021$ (9, 1) & $+0.083$ (13, 0) & $+0.124$ (15, 0) & $+0.129$ (15, 0) & $0.671$ \\
  Linear cluster & $+0.031$ (12, 2) & $+0.055$ (10, 3) & $+0.103$ (13, 0) & $+0.155$ (15, 0) & $0.553$ \\
  \end{tabular}
  \end{ruledtabular}
\end{table*}

\subsection{Mixed-state spectra and purity}\label{app_statedist}

For the first distribution shift of mixed input states, we sample an initial spectrum uniformly on the probability simplex and adjust it to a separately and uniformly drawn $u\in[0.55,1]$. To reach the drawn $u$, the initial spectrum is mixed with the uniform spectrum if its purity is too high, or with the pure-state spectrum concentrated on its largest component if its purity is too low. A Haar-random unitary sets the eigenvectors. The resulting spectra are not uniformly distributed over the spectra of a given purity. The second shift retains Eq.~(\ref{eq:dps}) but samples $u$ uniformly on $[0.30,0.55]$.

Over the $15$ conditions of each system size, the median differences from the better of MLE and SDP at $n=2$, $3$, $4$, and $5$ are $0.073$, $0.117$, $0.184$, and $0.223$ after lifting spectral degeneracy and $0.052$, $0.141$, $0.213$, and $0.304$ for the lower purity range, compared with $0.077$, $0.124$, $0.191$, and $0.229$ for the original distribution. The exceptions noted in Sec.~\ref{sec_result_shift} occur at $10^4$ shots and $r=100\%$: after lifting spectral degeneracy, the $n=2$ difference from the better of MLE and SDP is $-0.0036$ $[-0.0042,-0.0030]$, with the $95\%$ CI in brackets. For the lower purity range, the differences are $-0.00084$ $[-0.00094,-0.00074]$ at $n=2$ and $-0.00058$ $[-0.00100,-0.00017]$ at $n=3$.

\begin{figure*}[t]
  \centering
  \includegraphics[width=\textwidth]{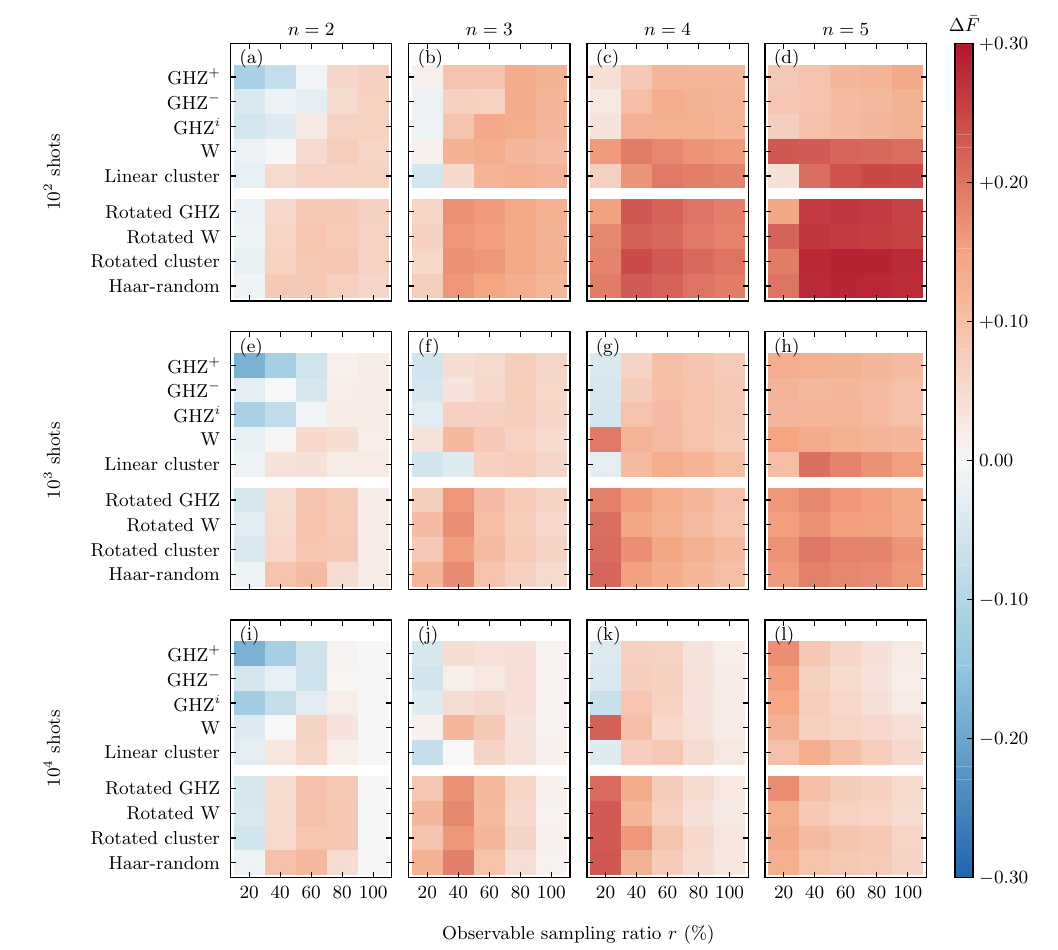}
  \caption{Fidelity differences $\Delta\bar F=\bar F^{(\text{Diff-QST})}-\max\{\bar F^{(\mathrm{MLE})},\bar F^{(\mathrm{SDP})}\}$ for structured states over all measurement conditions, using the networks for pure input states, on a common color scale; positive values favor \mbox{Diff-QST}. Panel rows correspond to $10^2$, $10^3$, and $10^4$ shots per observable and panel columns to $n=2,\ldots,5$; within each panel, columns are observable sampling ratios, the upper five rows are the canonical GHZ$^{+}$, GHZ$^{-}$, GHZ$^{i}$, W, and linear cluster states, and the lower four rows are the locally rotated GHZ, W, and linear-cluster ensembles and the Haar-random ensemble. Each canonical-state mean uses $100$ independent measurement acquisitions, and each ensemble mean $300$ states. CIs are not displayed; Table~\ref{tab:structured} summarizes them for the canonical states.}
  \label{fig:structured_heatmap}
\end{figure*}

\subsection{Readout miscalibration}\label{app_ood_summary}

Actual readout rates are multiplied by $s=0.5$ or $2$, while the assumed rates remain nominal. Over the $60$ conditions of each input-state ensemble, the median fidelity difference from the better of MLE and SDP is $0.1001$ at $s=0.5$ and $0.1427$ at $s=2$ for pure input states, and $0.1369$ and $0.1520$ for mixed input states. Targets and subsets are shared across scales. Doubling the readout-error rates gives median fidelity losses, relative to nominal readout, of $0.0108$, $0.0443$, and $0.0421$ for \mbox{Diff-QST}, MLE, and SDP, respectively, for pure input states, and $0.0203$, $0.0257$, and $0.0190$ for mixed input states. Rescaling the actual rates changes both the readout error in the data and its mismatch with the assumed model, so the test does not isolate calibration bias.

\subsection{Structured states}\label{app_structured}

The canonical GHZ states are $(|0\rangle^{\otimes n}+e^{i\phi}|1\rangle^{\otimes n})/\sqrt2$, with $\phi=0,\pi,\pi/2$ for GHZ$^+$, GHZ$^-$, and GHZ$^i$. The W state is the equal superposition of the $n$ computational-basis strings containing exactly one $1$. The linear cluster state is obtained by applying controlled-$Z$ gates between adjacent qubits of an open chain initialized in $|+\rangle^{\otimes n}$, where $|+\rangle=(|0\rangle+|1\rangle)/\sqrt2$.

Figure~\ref{fig:structured_heatmap} shows the condition-level fidelity difference $\Delta\bar F$ from the better of MLE and SDP for the canonical states and for the locally rotated and Haar-random ensembles. The canonical means average $100$ independent measurement acquisitions per condition, each drawing its own observable subset for $r<100\%$; the ensemble means average $300$ states, each with the three subsets shared by the ensemble. The CIs resample acquisitions for a canonical state and states for an ensemble, so the canonical CIs include subset variation, whereas the ensemble CIs are conditional on the three subsets.

Local rotations preserve ideal-input entanglement but can change target purity under the basis-dependent preparation channel. At $n=2$, GHZ, W, and linear cluster states are locally equivalent, so their rotated ensembles sample the same distribution; canonical representatives can differ under basis-dependent preparation noise.
For the locally rotated ensembles at $n=3,4,5$, median fidelity differences range from $0.108$ to $0.183$, with both CIs above zero in all $15$ conditions. Table~\ref{tab:structured} shows the more variable canonical-state results. At $n=2$, all five canonical representatives are maximally entangled and atypical of Haar-random states; the GHZ representatives have negative median differences, whereas W and linear cluster states have positive medians. The rotated ensembles have positive medians; at $r=20\%$, where both CIs lie below zero for every shot count (Sec.~\ref{sec_result_shift}), the differences are $-0.056$ to $-0.014$ and remain $-0.035$ to $-0.015$ after averaging over all $455$ three-observable subsets, with both CIs below zero for every ensemble and shot count.

\subsection{Weaker preparation noise}\label{app_channel}

All three preparation-channel parameters are scaled together by $s_{\mathrm p}=0.25$ or $0.5$, leaving the two ideal input-state ensembles unchanged. The target is the resulting prepared state at each strength. Over the $36$ conditions of each input-state ensemble, the median fidelity difference from the better of MLE and SDP is $0.0329$ at $s_{\mathrm p}=0.25$ and $0.0642$ at $s_{\mathrm p}=0.5$ for pure input states, and $0.0898$ and $0.1082$ for mixed input states. At the most negative difference, $\Delta\bar F=-0.1010$ (Sec.~\ref{sec_result_shift}), the mean fidelities are $0.7977$, $0.8987$, and $0.8937$ for \mbox{Diff-QST}, MLE, and SDP, respectively. These tests change the target distribution; they neither isolate purity from other state properties nor represent temporal fluctuations of the noise during one acquisition.

\clearpage

\bibliography{reference}

\end{document}